%% file: main.tex
\documentclass[12pt,onecolumn,twoside]{IEEEtran}
\IEEEoverridecommandlockouts

\usepackage{color}
\usepackage{cite}
\usepackage[subrefformat=parens]{subcaption}   
\usepackage{tabularx}     
\usepackage{textcomp}     
\usepackage{threeparttable} 
\usepackage{nth}          
\usepackage{gensymb}      
\usepackage{graphicx}     
\usepackage{graphics}
\usepackage{mathtools}    
\usepackage{amsmath}
\usepackage{amsfonts}
\usepackage{amsthm}
\usepackage{verbatim}
\usepackage{amssymb}
\usepackage{threeparttable}

\usepackage{stfloats}
\usepackage{setspace}
\usepackage{caption}
\usepackage{titlesec}
\usepackage{todonotes}
\usepackage{bbm}

\newcommand{\y}{\mathbf{y}}

\newcommand{\x}{\mathbf{x}}
\newcommand{\Sig}{\mathbf{S}}
\newcommand{\HM}{\mathbf{H}}
 
\usepackage[linesnumbered,vlined,ruled]{algorithm2e}

\begin{document}
\title{Learning-Based Near-Orthogonal Superposition Code for MIMO Short Message Transmission}

\author{Chenghong Bian, Chin-Wei Hsu, Changwoo Lee, Hun-Seok Kim,~\IEEEmembership{Senior Member,~IEEE} 
\thanks{The materials in this paper were presented in part at the IEEE International Conference on Communications (ICC), 2022 \cite{NOS}. This work was funded in part by NSF CAREER \#1942806.}
\thanks{C. Bian, C. Hsu, C. Lee and H. Kim are with the Department
of Electrical and Computer Engineering, University of Michigan, Ann Arbor,
MI, 48109 USA e-mail: (chbian@umich.edu; chinweih@umich.edu; cwoolee@umich.edu; hunseok@umich.edu)}

}


\maketitle \thispagestyle{empty}

\begin{abstract}
Massive machine type communication (mMTC) has attracted new coding schemes optimized for reliable short message transmission. In this paper, a novel deep learning-based near-orthogonal superposition (NOS) coding scheme is proposed to transmit short messages in  multiple-input multiple-output (MIMO) channels for mMTC applications. In the proposed MIMO-NOS scheme, a neural network-based encoder is optimized via end-to-end learning with a corresponding neural network-based detector/decoder in a superposition-based auto-encoder framework including a MIMO channel. The proposed MIMO-NOS encoder spreads the information bits to multiple near-orthogonal high dimensional vectors to be combined (superimposed) into a single vector and reshaped for the space-time transmission.  For the receiver, we propose a novel looped $K$-best tree-search algorithm with cyclic redundancy check (CRC) assistance to enhance the error correcting ability in the block-fading MIMO channel. Simulation results show the proposed MIMO-NOS scheme outperforms maximum likelihood (ML) MIMO detection combined with a polar code with CRC-assisted list decoding by 1 -- 2 dB in various MIMO systems for short (32 -- 64 bit) message transmission. 
\end{abstract}


\vspace{15mm}

\IEEEpeerreviewmaketitle

\input{introduction.tex}

\input{dl_model.tex}

\input{codewords.tex}

\input{kbest.tex}

\input{evaluation.tex}

\vspace{-1.0mm}
\section{Conclusion}
\vspace{-1.0mm}

This paper proposes a novel deep learning based MIMO-NOS coding scheme for reliable transmission of short messages in MIMO channels. The proposed end-to-end framework enables the encoder to successfully learn near-orthogonal superposition codewords with the aid of a neural network decoder. To improve the error rate performance, we propose and evaluate a CRC-assisted looped $K$-best decoder, which significantly outperforms the neural network decoder used during the training. We characterize the proposed MIMO-NOS coding and  provide empirical evaluation  with different MIMO settings and NOS encoding parameters. Simulation results show the proposed  MIMO-NOS scheme outperforms CRC-aided list decoding polar codes with maximum likelihood MIMO detection by 1 -- 2 dB in various MIMO configurations for short (32 -- 64 bits) message transmission.

\bibliographystyle{IEEEtran}
\bibliography{IEEEabrv,ref}

\end{document}

%% file: introduction.tex
\section{Introduction}


Massive machine type communication (mMTC) is an essential technology for next generation wireless standards to enable a wide range of applications including health, security and transportation \cite{mMTC1,mMTC2,mMTC3}. These applications, by nature, typically employ short messages/packets carrying a relatively small number of information bits, which make conventional codes designed with a long block length assumption less effective with relatively small error exponents and/or non-negligible coding gain losses. Polar codes with list decoding \cite{polarld} are proven to be more reliable compared with other modern codes such as LDPC and turbo codes for short block lengths \cite{compare_codes}. However, their performance is far from capacity, and thus new coding schemes have been actively investigated for short message transmission \cite{Coskun2019}. 

Hyper-dimensional modulation (HDM) is a recently proposed non-orthogonal modulation scheme for short packet communications \cite{HDM2018, HDM2019, HDMkbest}. HDM can be seen as a joint coding-modulation method and a special type of superposition codes \cite{SPARC2014}.
Instead of using a random codebook as in typical superposition codes, HDM uses fast Fourier transformation (FFT) and pseudo-random permutations to encode sparse pulse position modulated information vectors to a non-sparse superimposed hyper-dimensional vector for efficient encoding and decoding.
HDM was first proposed with a demodulation algorithm using an iterative parallel successive interference cancellation (SIC) technique \cite{HDM2018}. It is then extended using a $K$-best decoding algorithm \cite{HDMkbest} in AWGN and interference-limited channels to outperform the state-of-the-art CRC-assisted polar codes \cite{polarld} applied to binary phase-shift keying (BPSK) under the same spectral efficiency. Despite its excellent reliability and low complexity for short message packets, the hand-crafted encoding scheme using FFT and pseudo-random permutation for codeword generation in HDM is sub-optimal. It is shown in our prior work \cite{NOS} that a deep learning-based near-orthogonal superposition (NOS) encoding scheme can outperform HDM in single antenna additive white Gaussian noise (AWGN) channels.


In order to overcome the limitations of hand-crafted modulation and coding schemes, data-driven learning with deep neural networks (DNNs) has been applied to the realm of channel coding \cite{deepdec,deepldpc,deepae,LEARN,jiang2019turbo}. One of early applications of DNNs is to decode linear block codes replacing hand-crafted decoding algorithms for polar codes and LDPC with unmodified encoders. Taking advantage of powerful deep learning, prior schemes \cite{deepdec} and \cite{deepldpc} show improved decoding performance and enhanced robustness under various channel conditions. Meanwhile, new channel codes have been recently investigated via end-to-end learning. A DNN-based learned code was originally introduced in \cite{deepae} where the encoder learns a joint coding and modulation scheme generating a length-7 codeword from a length-16 one-hot input to achieve the performance similar to that of (7,4) Hamming code. 
The authors in \cite{LEARN} propose an RNN-based auto-encoder that emulates a  convolution code (CC) which takes the bit-sequence input instead of processing an one-hot encoded input vector. This \textit{learned} CC outperforms conventional CC to attain lower bit/packet error rates (BER/PER). In \cite{jiang2019turbo}, the authors propose a learned turbo auto-encoder which employs convolutional neural networks (CNNs) and interleaving. The decoder in \cite{jiang2019turbo} unfolds the iterative decoding process to multiple DNN layers to achieve the BER performance comparable to that of the conventional turbo code.

Meanwhile, researches are actively extending deep learning to multiple-input multiple-output (MIMO) detection problems. DetNet proposed in\cite{DetNet} unfolds the projected gradient descent algorithm via deep learning to achieve near optimal detection performance with significantly improved running speed. The authors in \cite{JMIMODD} further extend the topic to joint MIMO detection and polar decoding, where a DNN-based receiver takes both the received signal and the estimated channel state information (CSI) as the input to produce the estimated information sequence output. Their evaluation shows the DNN-based joint detection and decoding scheme outperforms the conventional iterative MIMO receiver where soft-decision information is exchanged between a maximum likelihood (ML) sphere decoder and a polar decoder to achieve near-optimal performance. However, their DNN-based receiver can only handle very short packets with 16 information bits and each MIMO channel configuration requires a specifically trained neural network model. These limitations make it rather impractical for emerging mMTC applications. 

\begin{figure}[t]
\centering
\includegraphics[width=0.9\linewidth]{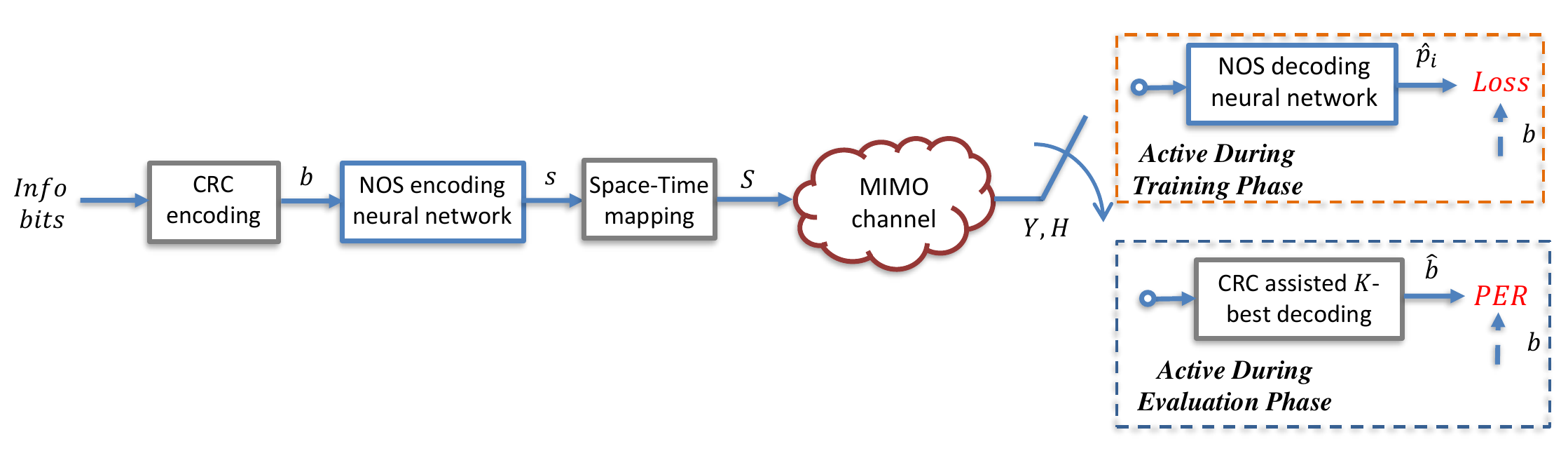}
\vspace{-0.3cm}
\caption{Encoding and decoding flow of the proposed MIMO-NOS scheme.}
\label{fig:flow_chart}
\end{figure}
Inspired by aforementioned HDM and deep learning-based coding, we originally introduced a DNN-based near-orthogonal superposition (NOS) coding scheme in \cite{NOS} to learn a near-orthogonal codebook for superimposed transmission of short packets in single-input single-output AWGN channels. In this paper, we further extend the NOS code to MIMO configurations constructing a \textit{learned MIMO-NOS coding scheme}. In our approach, an information bit sequence is first appended by cyclic redundancy check (CRC) bits to improve the reliability in low signal-to-noise ratio (SNR) scenarios. The CRC appended bit sequence $\mathbf{b}$ is transformed into several one-hot coded vectors which are fed into the MIMO-NOS encoder followed by a simple space-time (ST) coding block that maps the encoder output to different transmit antennas and time slots. To learn a good MIMO-NOS codebook, a DNN-based receiver that integrates a residual-assisted minimum mean square error (MMSE) MIMO equalizer/detector along with a neural decoder is jointly trained to enable end-to-end back-propagation through the encoder, MIMO channel model, and decoder. Upon learning a good MIMO-NOS codebook, we employ a CRC-assisted looped $K$-best tree-search decoding algorithm to improve the error rate performance beyond the limitation of the learned MIMO detector/decoder used for the training. The overall datapath of the proposed MIMO NOS scheme is shown in Fig.\ref{fig:flow_chart}.

The main contributions of this paper are summarized as follows:
\begin{enumerate}
    \item A novel deep learning-based near-orthogonal superposition code is proposed for reliable short packet transmission in MIMO channels. To the best of our knowledge, it is the first work that jointly learns the channel coding, modulation, MIMO detection, and channel decoding in an end-to-end fashion.
    \item A new CRC-assisted looped $K$-best decoding algorithm is designed to outperform the DNN-based receiver used during the training. The proposed decoding algorithm finds the top-$K$ bit-sequences maximizing the (approximated) posterior probability to significantly improve the PER performance beyond the capability of the DNN-based receiver utilized to learn the MIMO-NOS codebook. 
    \item Analysis on the learned MIMO-NOS codebook is provided to characterize the codebook properties, derive detection/decoding metrics for the looped $K$-best decoder, and study the performance of the proposed algorithm. It is also shown that the learned MIMO-NOS codebook can be applied to different MIMO configurations with robust performance via simple space-time mapping without retraining the encoder network. 
    \item Extensive numerical evaluations are performed to quantify the gain of the proposed learned MIMO-NOS scheme compared to the maximum likelihood (ML) MIMO detection with CRC-aided list decoding polar codes, which is one of the state-of-the-art baseline schemes. 
\end{enumerate}

Throughout the paper,  scalar variables are represented with normal-face letters $x$ while matrices and vectors with \textbf{upper} and \textbf{lower} case letters, $\mathbf{X}$ and $\mathbf{x}$, respectively. Transpose and Hermitian operators are denoted by $(\cdot)^T$, $(\cdot)'$, respectively. Moreover, $\Re(x)$ ($\Im(x)$) denotes the real (imaginary) part of $x$, $Diag(\mathbf{X})$ denotes the diagonal elements of matrix $\mathbf{X}$, and $vec(\mathbf{X})$ transforms the matrix $\mathbf{X}$ into a column vector. Finally, we denote the Frobenius norm of matrix $\mathbf{X}$ as $||\mathbf{X}||_F$.

%% file: dl_model.tex
\section{MIMO-NOS Code Learning}
In this section, we briefly recap the conventional coded MIMO transceiver for the baseline, and then introduce a neural network structure for the proposed MIMO-NOS scheme and its training methodology.

\subsection{Conventional MIMO Transceiver}
Let $\mathbf{b}$ and $\mathbf{c}$ denote the information bit sequence and corresponding coded bit sequence with length $N_t M_c log_2Q$ bits. The coded sequence $\mathbf{c}$ is mapped to a matrix $\mathbf{S}$ of dimension $N_t \times M_c$ whose entries are chosen from a complex constellation set (e.g., QPSK) with $Q$ symbols. $N_t$ is the number of transmit antennas and $M_c$ is the number of MIMO channel use for transmission. The received signal $\mathbf{Y} \in \mathbb{C}^{N_r\times M_c}$ with $N_r$ receive antennas can be written as:
\begin{align}
\mathbf{Y} = \HM \mathbf{S} +\mathbf{N},
\label{eq:mimo_detect}
\end{align}
where $\HM \in \mathbb{C}^{N_r\times N_t} $  is the complex MIMO channel which is assumed to be perfectly known to the receiver and $\mathbf{N}$ is the complex Gaussian noise whose entries are i.i.d. with zero mean and element-wise variance $\sigma^2$. In this paper, we assume each element of $\HM$ is an i.i.d. complex Gaussian random variable with zero mean and unit variance. $\HM$ is randomly realized for each and every packet.

There are numerous MIMO detection algorithms to  solve \eqref{eq:mimo_detect} and obtain  soft decisions of bits in the sequence $\mathbf{S}$ with a simplifying assumption that  bits in $\mathbf{c}$ are independent. With that assumption, the ML MIMO detector is the optimal scheme\footnote{The complexity of ML detection may be prohibitive for a large MIMO system with a high constellation. Low complexity close-to-ML algorithms are available but they are beyond our consideration for this paper.}, and thus it is applied to the baseline as briefly introduced in the following.

Consider the log-likelihood ratio (LLR) $L$ for a certain bit $\mathbf{c}_k$ from $\mathbf{c}$ given $\mathbf{y} = \HM \mathbf{s} +\mathbf{n}$ where $\mathbf{s}$ is a $N_t \times 1$ transmit vector (a column of $\mathbf{S}$ that involves $\mathbf{c}_k$), and $\mathbf{y}$ and $\mathbf{n}$ are corresponding received and noise vectors, respectively. The LLR of $\mathbf{c}_k$ can be obtained by
\begin{align}
L(\mathbf{c}_k|\mathbf{y},\HM) = \ln \frac{P[\mathbf{c}_k=+1|\mathbf{y},\HM]}{P[\mathbf{c}_k=-1|\mathbf{y},\HM]}.
\label{eq:llr_val}
\end{align}
Applying Bayes's rule and assuming equal probability of the bit symbols, the $L$ values are obtained by
\begin{align}
    \label{eq:final_llr} &L(\mathbf{c}_k|\mathbf{y},\HM) = \ln \frac{\sum_{\mathbf{x}\in \mathbf{X}_{k,+1}}\exp\{\frac{-||\mathbf{y}-\HM \mathbf{x}||_2^2}{2\sigma^2}\}}
    {\sum_{\mathbf{x}\in \mathbf{X}_{k,-1}}\exp\{\frac{-||\mathbf{y}-\HM \mathbf{x}||_2^2}{2\sigma^2}\}}
\end{align}
where each set $\mathbf{X}_{k,+1}=\{\mathbf{x}|\mathbf{c}_k=+1\}$ or $\mathbf{X}_{k,-1}=\{\mathbf{x}|\mathbf{c}_k=-1\}$ contains $2^{(N_t log_2Q)-1}$ bit sequences of length $N_tlog_2Q$ bits, enumerating all possible bit sequences given $\mathbf{c}_k=1$ or $-1$. 
In the baseline scheme, we first calculate the LLR (i.e., soft decision) for each coded bit using  \eqref{eq:final_llr}, and then we feed them into the subsequent soft-input channel decoder (such as a CRC-assisted list polar decoder) to recover the original information bit sequence $\textbf{b}$.

\vspace{-2mm}
\subsection{MIMO-NOS Coding}

\begin{figure*}[ht]
\centering
\includegraphics[width=0.9\linewidth,height=0.22\linewidth]{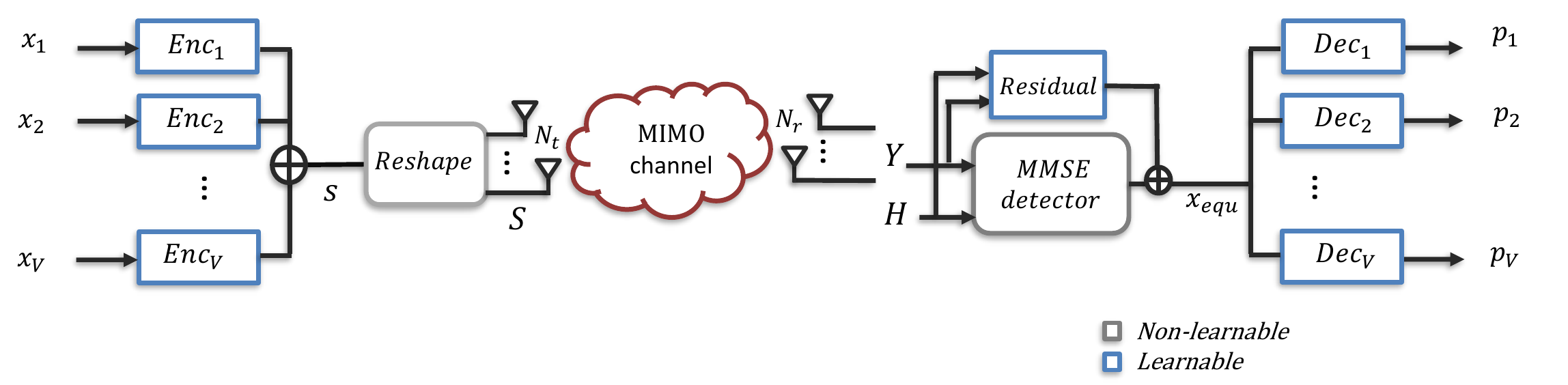}
\caption{The proposed neural network-based MIMO-NOS encoding and decoding structure for training.}
\label{fig:network}
\end{figure*}

We now discuss the proposed deep learning based MIMO-NOS scheme. Consider a sequence of information bits $\mathbf{b}$ whose length is $V\times m$ bits. It is split into $V$ smaller bit sequences $\mathbf{b}_v, v = 1,\cdots,V$ each carrying $m$ bits. Each $\mathbf{b}_vi$ is converted to an one-hot vector $\mathbf{x}_v$ with length $M = 2^m$ whose only non-zero position (with value 1) is determined by $\mathbf{b}_v$. 
A superposition code is defined by a complex-valued codebook $\cal C$ with dimension $(V, D, M)$ where $D$ is the codeword length. The codeword corresponding to the bit sequence $\mathbf{b}_v$ is obtained by ${\cal C}[v,:,:]\x_v $ (i.e., matrix-vector multiplication), whose dimension is $D\times 1$. The superimposed transmit vector $\mathbf{s}$ with length $D$ for the entire bit sequence $\mathbf{b}$ is then obtained by adding (superimposing) $V$ codewords such that:
\vspace{-2mm}
\begin{align}
\mathbf{s} = \sum_{v=1}^V {\cal C}[v,:,:]\x_v .
\end{align}

Conventional superposition codes adopt pseudo-random codebooks $\cal C$, e.g.,  random (complex) Gaussian codebooks as in \cite{SPARC2014}, whereas a more efficient scheme such as HDM \cite{HDM2018} defines the codebook using the discrete (fast) Fourier transform (DFT/FFT) matrix along with pseudo-random permutations. There exist efficient decoding algorithms for these schemes in the AWGN channel including successive interference canceling (SIC) \cite{HDM2018,SPARC2014} and approximate message passing (AMP) \cite{SPARC_AMP} schemes. Although these superposition codes are proven to be capacity achieving when the block length goes to infinity \cite{SPARC2014}, a pseudo-random codebook is shown to be less effective under short block lengths \cite{Finite_SPARC} . Thus, a new near-orthogonal superposition (NOS) code is proposed in our prior work \cite{NOS} with a learned codebook $\cal C$ that consists of near-orthogonal codewords obtained via deep learning. In \cite{NOS}, a tree-based decoding algorithm for the learned codebook is proposed to outperform hand-crafted superposition codes such as HDM \cite{HDM2018} as well as BPSK transmission with CRC-aided list polar codes in the AWGN (non-MIMO) channel. However, the learned codebook and the decoding scheme proposed in \cite{NOS} are not directly applicable to MIMO channels. Hence, we propose an extended MIMO-NOS scheme for short message MIMO transmission as follows.

Fig. \ref{fig:network} shows the overview of the proposed MIMO-NOS transmission scheme. Multiple ($V$) one-hot vectors $\mathbf{x}_v$ are fed to dedicated neural network-based encoders $Enc_v$ to generate real-valued coded vectors $\tilde{\mathbf{s}}_v=Enc_v(\x_v)$ of length $D$. Each $Enc_v, v\in [1,V]$ has the same neural network structure that consists of linear layers, batch normalization layers, and non-linear activation functions. Since each $\tilde{\mathbf{s}}_v$ conveys the same amount of information, we assign the same energy $\tilde{\mathbf{s}}'_v\tilde{\mathbf{s}}_v=\frac{D}{2V}$ to each $\tilde{\mathbf{s}}_v$ using a power normalization layer at the end of each $Enc_v$. Instead of transmitting a real-valued signal, we convert the length-$D$ real-valued vector $\tilde{\mathbf{s}}_v$ into a complex vector $\mathbf{s}_v$, denoted as $\mathbf{s}_v = Complex(\tilde{\mathbf{s}}_v)$, to improve spectral efficiency by taking the first $D/2$ (assume $D$ is even) elements of $\tilde{\mathbf{s}}_v$ as the real part and the rest as the imaginary part. The superimposed signal $\mathbf{s}$ is obtained by adding  all $\mathbf{s}_v, v=1, 2, ... ,V$:
\begin{align}
\mathbf{s} = \sum_{v=1}^V Complex(Enc_v(\x_v)).
\end{align}

Then, we map $\mathbf{s}$ to different transmit antennas and time slots for space-time coding, which is extensively studied in \cite{STT,STBC}. In this paper, we adopt simple reshaping  that converts (reshapes) $\mathbf{s}$ to a transmit block $\mathbf{S} = Reshape(\mathbf{s}), \mathbf{S} \in \mathbb{C}^{N_t\times M_c}$, where $N_t$ is the number of transmit antennas, $M_c$ is the number of channel uses (or time slots), and $N_t M_c=D/2$ holds.

We assume the block-fading (quasi-static) MIMO channel, where channel coefficients in $\mathbf{H}\in \mathbb{C}^{N_r\times N_t}$ are instantiated as i.i.d. complex Gaussian random variables that remain constant for a single block transmission ($M_c$ channel uses). The next block observes an independent random channel realization $\mathbf{H}$ following the same model used for the conventional MIMO transmission. The received signal $\mathbf{Y}$ after  the MIMO channel can be expressed as $\mathbf{Y} = \mathbf{H} \mathbf{S} + \mathbf{N}$ as in \eqref{eq:mimo_detect} where $\mathbf{N}\in \mathbb{C}^{N_r\times M_c}$ is the complex Gaussian noise whose entries are i.i.d. with zero mean and element-wise variance $\sigma^2$. The signal to noise ratio (SNR) of the system is defined as:
\begin{align}
SNR = \frac{\mathbb{E}(||\mathbf{H} \mathbf{S}||^2_F)}{N_t \mathbb{E}(||\mathbf{N}||_F^2)} = \frac{N_r D/2}{N_t N_r M_c \sigma^2} = \frac{1}{\sigma^2},
\label{eq:snr_def}
\end{align}
where we use the fact that $N_t M_c=D/2$ and $\mathbb{E}(||\mathbf{S}||^2_F) = D/2$ ($\mathbf{s}_v$ are near-orthogonal to each other as examined in the later section). Note that, in practical systems, $\mathbf{H}$ can be obtained at the receiver by applying channel estimation algorithms \cite{MIMO_CHEST1,MIMO_CHEST2,MIMO_CHEST3} to the received pilots. However, we make a simplifying assumption that $\mathbf{H}$ is perfectly available at the receiver throughout this paper.

\subsection{Learned MIMO-NOS Receiver}
To learn a set of MIMO-NOS encoders $Enc_v, v\in [1,V]$, the training process uses a matching set of MIMO-NOS decoders. For decoding, the received signal $\mathbf{Y}$ and the MIMO channel $\HM$ are first fed to the residual-assisted MIMO detector/equalizer that consists of a conventional MMSE equalization module which serves as the backbone and a residual connection neural network module to compensate the output of the MMSE equalization module as shown in Fig. \ref{fig:network}. This residual-assisted structure is inspired by \cite{JSCC_OFDM} and it outperforms the MMSE-only structure as well as the neural network-only structure. 

The MMSE equalization module output is:
\begin{align}
\mathbf{X}_{MMSE} &= (\HM' \HM+\frac{\sigma^2}{P}\mathbf{I}_{N_t})^{-1} \HM ' \mathbf{Y},
\label{eq:MMSE_EQU}
\end{align}
where $\sigma^2$ and $P$ denotes the noise power and signal power, respectively. 
The residual module, denoted by $Res$, is a neural network that takes the real-valued received signal $\tilde{\mathbf{Y}} \in \mathbb{R}^{2N_r\times M_c}$ and the real-valued vectorized channel state information (CSI), $\tilde{\mathbf{h}}\in \mathbb{R}^{2N_tN_r\times 1}$ as the input. Note that $\tilde{\mathbf{Y}}$ and $\tilde{\mathbf{h}}$ are obtained by concatenating (vectorizing) real and imaginary parts of $\mathbf{Y}$ and $\mathbf{H}$, respectively.  Further, we concatenate $\tilde{\mathbf{h}}$ to each column of $\tilde{\mathbf{Y}}$ to form a larger matrix, defined as $Cat(\tilde{\mathbf{Y}},\tilde{\mathbf{h}}) \in \mathbb{R}^{G\times M_c}$ with $G = 2N_r(N_t+1)$. Then $Res$ module processes this concatenated signal to produce the final output of the residual-assisted MIMO detector expressed by:
\begin{align}
\tilde{\mathbf{X}}_{EQU} &= \tilde{\mathbf{X}}_{MMSE} + Res(Cat(\tilde{\mathbf{Y}},\tilde{\mathbf{h}})),
\label{eq:RES_NET}
\end{align}
where $\tilde{\mathbf{X}}_{MMSE}\in \mathbb{R}^{2N_t\times M_c}$ is the real-valued version (via concatenating real and imaginary parts) of the MMSE equalization module output.

We then vectorize $\tilde{\mathbf{X}}_{EQU}$ to $\tilde{\mathbf{x}}_{equ}$ with $D$-dimension, which is fed into each decoder $Dec_v, v\in [1,V]$ as shown in Fig. \ref{fig:network}. Similar to $Enc_v$, these $Dec_v$ are neural networks that consist of linear layers, batch normalization layers, and non-linear activation functions.  Each $Dec_v$ for a specific $v$ is trained to produce/estimate the probability vector $\mathbf{p}_v = Dec_v(\tilde{\x}_{equ})$. The length of $\mathbf{p}_v$ is $M$ and $\mathbf{p}_v[m]$ represents the probability of $\x_v[m]=1$.

\begin{figure}[ht]
\centering
\includegraphics[width=0.8\linewidth]{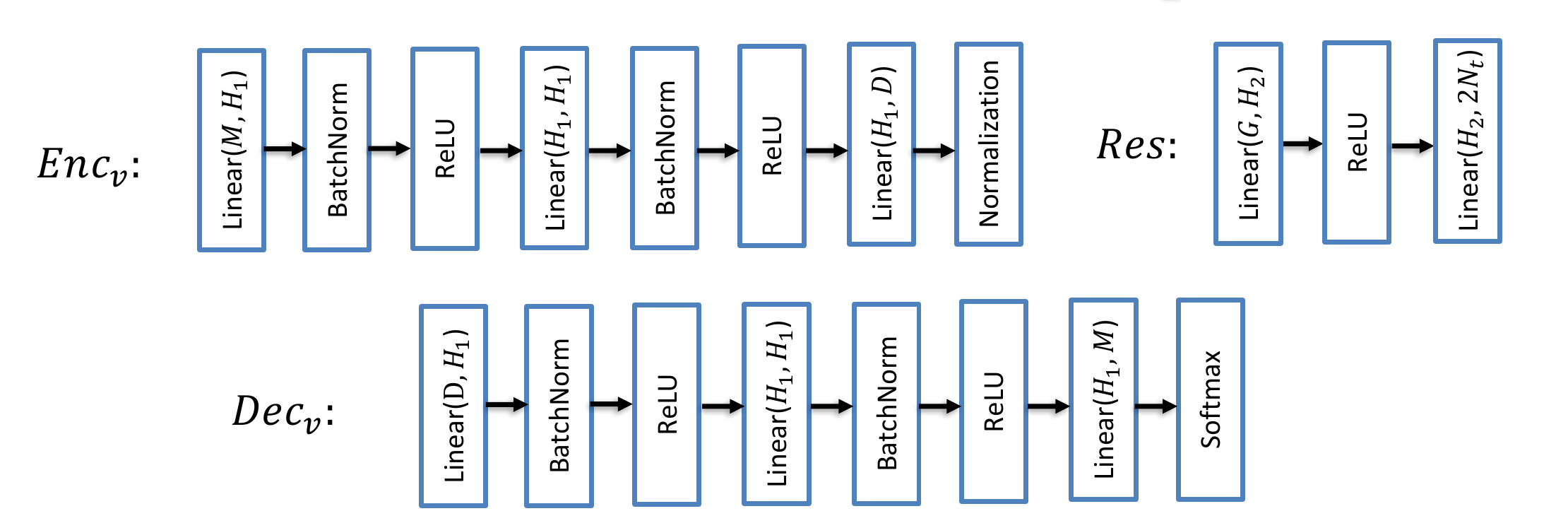}
\caption{Structures of each encoder decoder pair $Enc_v$ and $Dec_v, v\in [1,V]$ and the residual net $Res$. $H_1$ and $H_2$ denotes the number of hidden neurons in the model.}
\label{fig:detail_net}
\end{figure}

The detailed neural network structures of $Enc_v, Dec_v,$ and $Res$ are shown in Fig. \ref{fig:detail_net} where $H_1, H_2$ denote the number of neurons in the hidden layers. The training of $Enc_v, Dec_v,$ and $Res$ is performed by optimizing the  cross-entropy loss for each pair of the one-hot input $\mathbf{x}_v$ and the probability vector $\mathbf{p}_v$. Since $\mathbf{x}_v$'s assigned to different $Enc_v$'s are independent of each other, the total loss is the summation of pairwise losses:
\begin{align}
loss = -\sum_{v=1}^V {\sum_{m=1}^{M}{\mathbf{x}_v[m] \log(\mathbf{p}_v[m])}}.
\label{eq:loss}
\end{align}
The ADAM optimizer is used to train the proposed networks.

%% file: codewords.tex
\section{The Learned MIMO-NOS Codebook Properties}

In this section, we inspect the properties of the learned MIMO-NOS codebook before and after the MIMO channel.

\subsection{Codebook Properties Before the MIMO Channel}
The learned complex-valued codebook ${\cal C}$ with dimension $(V, D/2, M)$ is obtained by enumerating all length-$M$ one-hot vectors for each encoder after successful training:
\begin{align}
{\cal C}[v,:,m] = Complex(Enc_v(\x_m)).
\label{eq:codebook}
\end{align}

Similar to the analysis in \cite{NOS} for single antenna channels, we first analyze the properties of the constructed codebook ${\cal C}$ by observing the absolute values of inner products between codewords belonging to different encoders. This forms a cross-correlation tensor $c_{inter}$ with dimension $(V,V-1,M,M)$ which is defined as:
\begin{equation}
    \begin{split}
    c_{inter}[i,j,k,l] &= \frac{|\Re({\cal C}'[i,:,k] {\cal C}[j,:,l])|}{D/2V} \\ 
    i,j \in [1,V]; & i \neq j ; k, l \in [1,M].
    \end{split}
    \label{eq:inter_corr}
\end{equation}
Note that $c_{inter}$ quantifies the level of interference from the codewords belonging to different encoders, thus it represents the \textit{inter}-correlation property of the codebook. We further evaluate inner products between the codewords belonging to the \textit{same} encoder to form another tensor $c_{intra}$ with dimension $(V,M,M-1)$ defined as:
\begin{equation}
    \begin{split}
    c_{intra}[i,k,l] &= \frac{\Re({\cal C}'[i,:,k] {\cal C}[i,:,l])}{D/2V} \\ 
    i \in [1,V]; & k \neq l ; k, l \in [1,M],
    \end{split}
    \label{eq:intra}
\end{equation}
which represents the \textit{intra}-correlation property. Since the power of codewords are normalized, $c_{intra}$ directly reflects the L2-distance between codewords belonging to the same encoder. A small (or negative) $c_{intra}$ entry implies longer L2-distance for the corresponding pair, which is desirable to lower the error rate. Since the error performance of a code is mainly determined by its minimum distance, we are interested in the distribution of  entries of $c_{intra}$ with relatively large positive values.

\begin{figure}[t]
	\centering
	\begin{subfigure}{0.49\linewidth}
		\centering
		\includegraphics[width=0.9\linewidth]{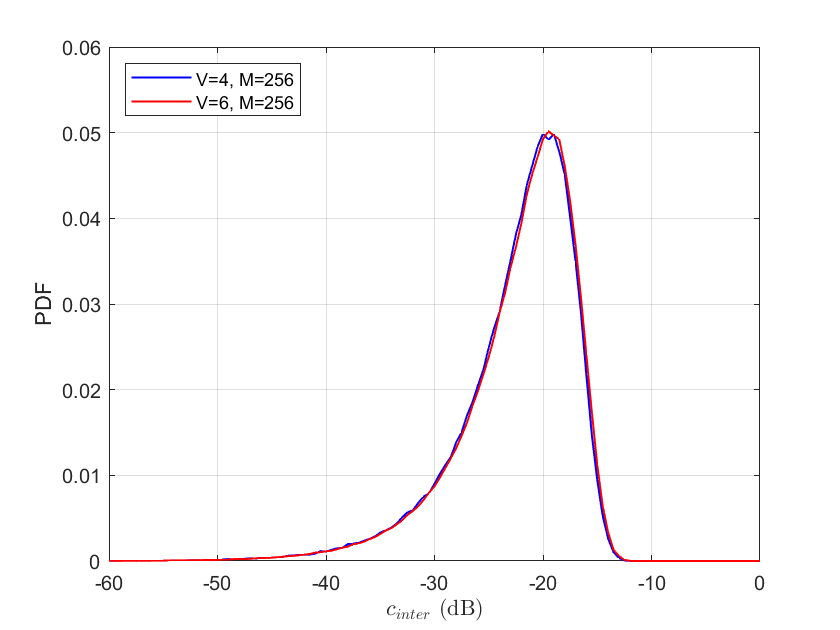}
		\caption{\textit{Inter}-correlation distribution before MIMO channels}
	\end{subfigure}
	\begin{subfigure}{0.49\linewidth}
		\centering
		\includegraphics[width=0.9\linewidth]{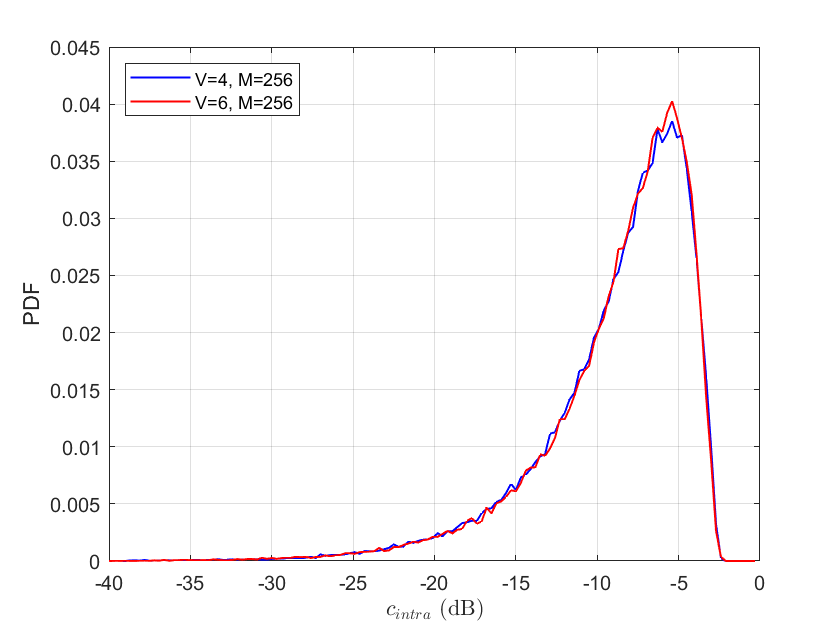}
		\caption{\textit{Intra}-correlation distribution before MIMO channels}
	\end{subfigure}
	\caption{The distribution of the absolute value of entries (positive entries) in $c_{inter}$ / $c_{intra}$ for the two codebooks learned with $(V = 4, M = 256, D = 64, N_t = N_r = 2)$ and $(V = 6, M = 256, D = 96, N_t = N_r = 2)$.}
	\label{fig:inter_corr}
\end{figure}

Fig. \ref{fig:inter_corr} (a) shows the distribution of entries (in dB) in $c_{inter}$ for the codebooks trained with $(V = 4, M = 256, D = 64, N_t = N_r = 2)$  and $(V = 6, M = 256, D = 96, N_t = N_r = 2)$. We observe that cross-correlation values are at least $\approx12$dB lower than the energy of each codeword ($D/2V$). This confirms that learned codewords belonging to different encoders are \textit{nearly-orthogonal} to each other. Similarly, Fig. \ref{fig:inter_corr} (b) shows the distribution of the positive entries (in dB) of $c_{intra}$ \eqref{eq:intra} for the same codebooks plotted in Fig. \ref{fig:inter_corr} (a). Note that the largest positive entry of $c_{intra}$ is $\approx2.5$dB lower than the energy of a codeword  implying that the minimum L2-distance among the codewords from the same encoder is not insignificant.

\subsection{Codebook Properties After a MIMO Channnel}
The MIMO-NOS codebook ${\cal C}$ exhibits the near-orthogonal property and reasonable minimum distances before MIMO transmission. This observation aligns with the results in \cite{NOS}, which only considers single antenna transmission cases. In this section, we further inspect codebook properties after the MIMO channel.

For a random MIMO channel realization, $\mathbf{H}$, whose entries are independent zero-mean complex Gaussian with unit variance, the post-channel codebook is updated from  \eqref{eq:codebook} to
\begin{align}
{\cal C}_{\mathbf{H}}[v,:,m] = vec(\HM \, Reshape({\cal C}[v,:,m]))
\label{eq:Hcodebook}
\end{align}
where ${\cal C}_{\mathbf{H}}\in \mathbb{C}^{V \times N_r M_c\times M}$. Following the same principle in \eqref{eq:inter_corr}, the \textit{inter}-correlation tensor $c_{inter}^{\mathbf{H}}\in \mathbb{R}^{V\times (V-1)\times M\times M}$ is obtained by:
\begin{equation}
    \begin{split}
    c_{inter}^{\HM}[i,j,k,l] &= \frac{|\Re({\cal C}'_{\HM}[i,:,k] {\cal C}_{\HM}[j,:,l])|}{N_r D/2V} \\ 
    i,j \in [1,V]; & i \neq j ; k, l \in [1,M],
    \end{split}
    \label{eq:H_corr}
\end{equation}
where the denominator $N_r D/2V$ is obtained by the expectation of $||{\cal C}_{\HM}[v,:,m]||_2^2$ over random realizations of $\HM$:
\begin{align}
\mathbb{E}(||{\cal C}_{\HM}[v,:,m]||_2^2) = {\cal C}'[v,:,m]\mathbb{E}((\mathbf{I}_{M_c} \otimes \HM)'(\mathbf{I}_{M_c} \otimes \HM)){\cal C}[v,:,m] = N_r D/2V.
\label{eq:rec_power}
\end{align}
In \eqref{eq:rec_power}, $\mathbf{I}_{M_c}$ is an $M_c \times M_c$ identity matrix,  $\otimes$ is the Kronecker product and $\mathbb{E}(\HM'\HM)=N_r\mathbf{I}_{N_t}$ holds. Similarly the \textit{intra}-correlation, $c_{intra}^{\HM}$ after the MIMO channel is defined by:
\begin{equation}
    \begin{split}
    c_{intra}^{\HM}[i,k,l] &= \frac{\Re({\cal C}'_{\HM}[i,:,k] {\cal C}_{\HM}[i,:,l])}{N_r D/2V} \\ 
    i \in [1,V]; & k \neq l ; k, l \in [1,M].
    \end{split}
    \label{eq:H_intra}
\end{equation}

\begin{figure}[t]
	\centering
	\begin{subfigure}{0.49\linewidth}
		\centering
		\includegraphics[width=0.9\linewidth]{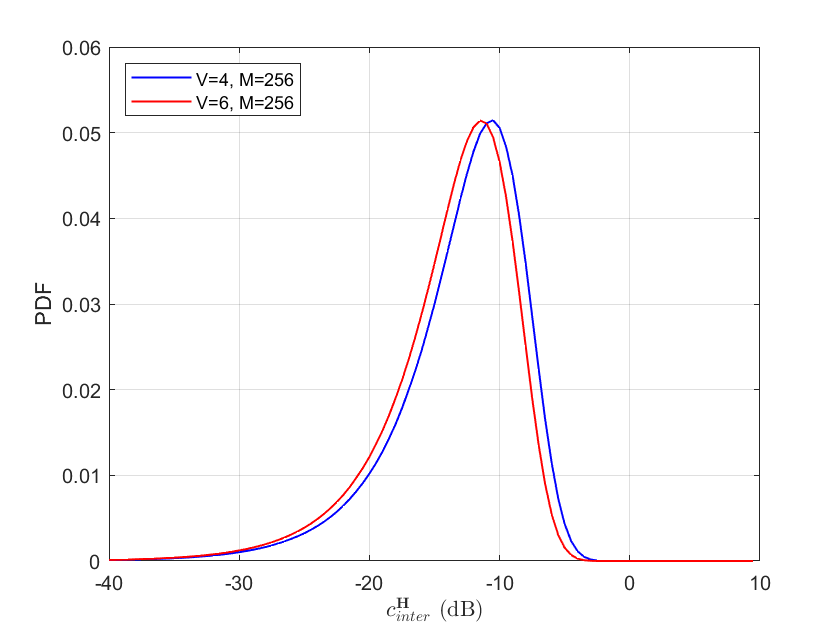}
		\caption{\textit{Inter}-correlation distribution after MIMO Channels}
	\end{subfigure}
	\begin{subfigure}{0.49\linewidth}
		\centering
		\includegraphics[width=0.9\linewidth]{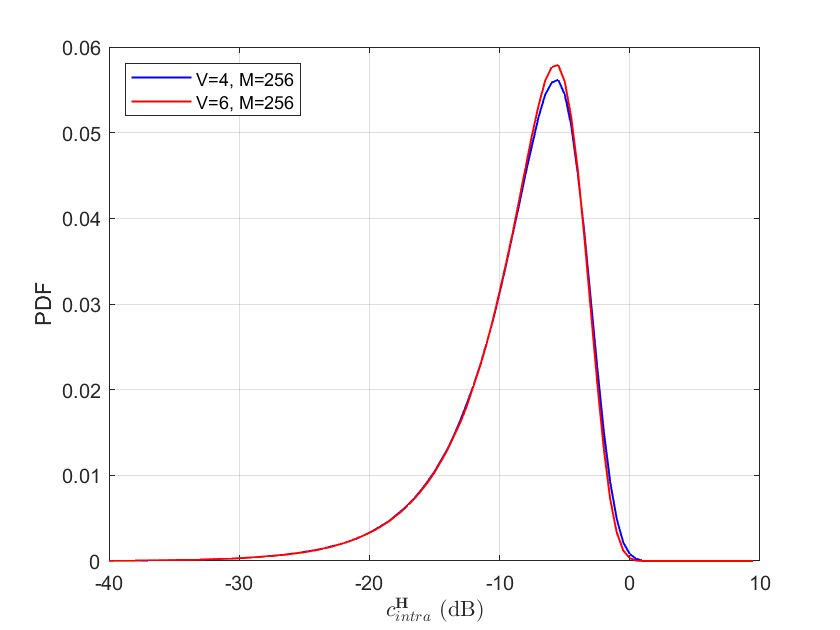}
		\caption{\textit{Intra}-correlation distribution after MIMO Channels}
	\end{subfigure}
	\caption{The distribution of the absolute values of $c_{inter}^{\mathbf{H}}$ entries and positive entries of $c_{intra}^{\mathbf{H}}$ using random channel realizations $\HM$ in a $4 \times 4$ MIMO system.}
	\label{fig:H_corr}
\end{figure}

To obtain empirical distributions, we randomly instantiate one thousand  $4\times 4$ MIMO channel matrix $\HM$'s and evaluate both $c_{inter}^\mathbf{H}$ and $c_{intra}^\mathbf{H}$ realizations. The distribution of absolute values of  $c_{inter}^\mathbf{H}$ entries  and the positive entries of $c_{intra}^{\mathbf{H}}$ are  plotted in Fig. \ref{fig:H_corr} (a) and (b) respectively. The codebooks used for this evaluation are the same ones used in Fig.  \ref{fig:inter_corr}. Fig. \ref{fig:H_corr} (a) shows that the correlation between codewords belonging to different encoders after a MIMO channel is not negligible, and thus they do not preserve the \textit{near-orthogonal} property any more. The maximum cross-correlation $c_{inter}^{\HM}$ is only $5dB$ lower than the expected energy of each received codeword ($N_r D/2V$) which is significantly higher than the maximum of the pre-channel correlation $c_{inter}$ shown in Fig. \ref{fig:inter_corr} (a). Meanwhile, $c_{intra}^{\mathbf{H}}$ shown in Fig. \ref{fig:H_corr} (b) has the largest positive element comparable to the expected codeword energy, implying significant minimum distance reduction between codewords from the same encoder after the MIMO channel.

%% file: kbest.tex
\vspace{2mm}
\section{$K$-best Assisted Decoding}

Significant post-channel interference and codeword distance reduction observed in the previous section motivate the need for an efficient algorithm to mitigate these issues to attain close-to-ML decoding performance. Since the ML solution is practically infeasible due to excessive complexity, we propose and investigate a new practical CRC-assisted $K$-best tree-search algorithm for the learned MIMO-NOS code. Later, we will show that the proposed algorithm significantly outperforms the neural network-based decoder which was used to train the learned-NOS codebook. 

\subsection{$K$-best MIMO-NOS Decoding}
The encoder and decoder neural network pair introduced in Section II is trained to minimize the number of bit errors per vector/codeword. However, typical mMTC applications do not tolerate any bit errors in a short packet, hence the primary objective of our scheme is to minimize the PER. For that, we include CRC bits in the information message to enhance the reliability of short packets in the low SNR regime. In our scenario, each transmitted block $\Sig \in \mathbb{C}^{N_t\times M_c}$ corresponds to a packet (which is obtained by space-time reshaping of a codeword, $\Sig = Reshape(\mathbf{s})$).

Consider the joint probability $P(\mathbf{x}_1^{m_1},\cdots, \mathbf{x}_V^{m_V}|\mathbf{Y},\HM)$ where $m_v \in [1,M]$. We desire to find the top-$K$ ($K$-best) combinations that maximize the joint probability over all possible combinations of one-hot vectors $\{\mathbf{x}_1^{m_1}, \cdots, \mathbf{x}_V^{m_V}\}$. Note that in \cite{NOS}, we have solved the top-$K$ searching problem of the learned NOS code in the single-input single-output (SISO) transmission AWGN channel. However, the assumption that the codewords are near-orthogonal \textit{after} the channel is no longer valid for the MIMO transmission as shown in Fig. \ref{fig:H_corr}. Thus the joint probability does not factorize into products of marginal probabilities $\prod_{v=1}^V p(\x_v|\mathbf{Y},\HM)$ as in the SISO AWGN channel case \cite{NOS}. For the MIMO-NOS code, a procedure of finding the top-$K$ combinations is proposed as follows.

\begin{figure}[t]
\centering
\includegraphics[width=0.8\linewidth]{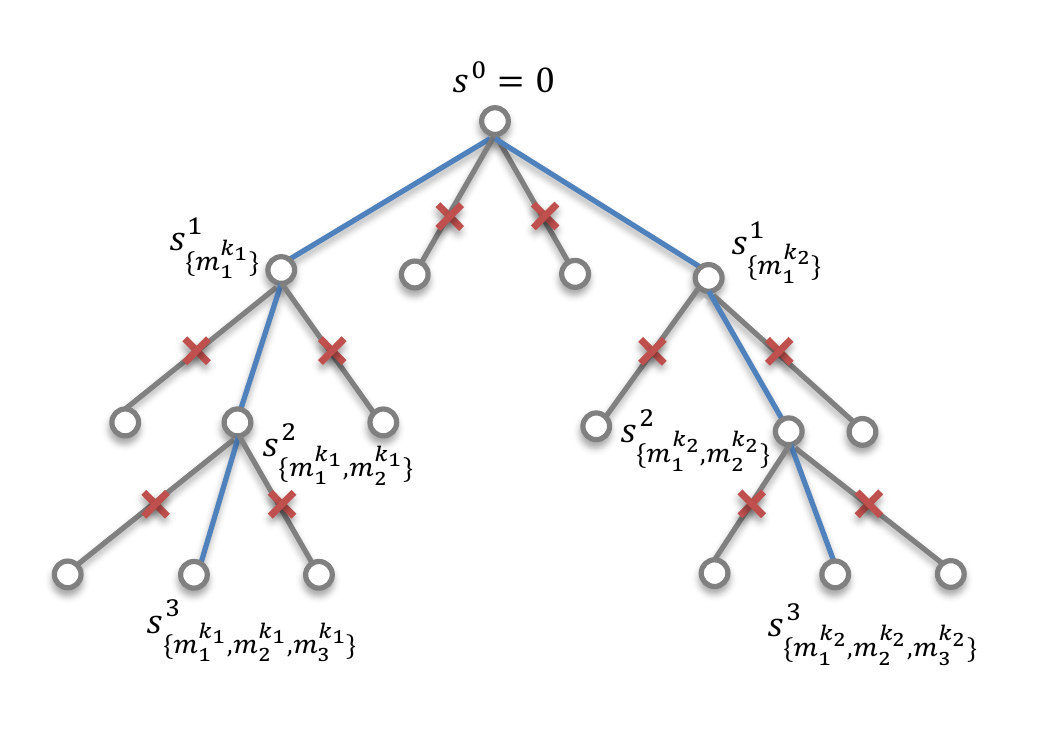}
\vspace{-0.3cm}
\caption{The proposed $K$-best algorithm. The two blue branches indicate $K=2$ survived paths in the tree.}
\label{fig:kbest}
\end{figure}

The joint probability of an NOS code follows the expression:
\begin{align}
P(\mathbf{x}_1^{m_1}&,\cdots, \mathbf{x}_V^{m_V}|\mathbf{Y},\HM) \propto \exp\{-\frac{1}{2\sigma^2} ||\mathbf{y}-\sum_{v=1}^V{{\cal C}_{\mathbf{H}}[v,:,m_v]}||^2_2\},
\label{eq:joint prob}
\end{align} 
where $\y \in \mathbb{C}^{N_r M_c\times 1}$ is the vectorized version of the received matrix $\mathbf{Y}$ and ${\cal C}_{\mathbf{H}}$ is the codebook corresponding to $\HM$ defined in \eqref{eq:Hcodebook}. The problem of finding $K$ candidates maximizing the joint probability is equivalent to finding $m_v$'s that minimize the L2-distance $||\mathbf{y}-\sum_{v=1}^V{{\cal C}_{\mathbf{H}}[v,:,m_v]}||^2_2$. It is practically infeasible for large $V$ and $M$ to identify the exact $K$-best candidates. To adopt the principle of $K$-best tree searching and pruning algorithms designed for near-ML MIMO detection \cite{kbest,kbest2}, we decompose the L2 term  in \eqref{eq:joint prob} into four terms: 
\begin{align}
||\mathbf{y}&-\sum_{v=1}^V{{\cal C}_{\mathbf{H}}[v,:,m_v]}||^2_2  =  ||\mathbf{y}-\sum_{v=1}^{V-1}{{\cal C}_{\mathbf{H}}[v,:,m_v]}||^2_2+||{\cal C}_{\mathbf{H}}[V,:,m_V]||_2^2 \notag  \\ &+2\Re({\cal C}'_{\mathbf{H}}[V,:,m_V]\sum_{v=1}^{V-1}{{\cal C}_{\mathbf{H}}[v,:,m_v]}- {\cal C}'_{\mathbf{H}}[V,:,m_V] \mathbf{y}),
\label{eq:l2_dist}
\end{align} 
where the first term is the same as the LHS except the summation is from 1 to $V-1$. To allow recursive metric evaluation, define the \textit{score metric} $s^{(l)} = ||\mathbf{y}-\sum_{i=1}^l{{\cal C}_{\mathbf{H}}[i,:,m_i]}||^2_2$, which can be expressed as:
\begin{align}
  s^{(l)} = s^{(l-1)}  + ||{\cal C}_{\mathbf{H}}[l,:,m_l]||_2^2 
  + 2\Re({\cal C}'_{\mathbf{H}}[l,:,m_l]\mathbf{u}^{(l-1)}- {\cal C}'_{\mathbf{H}}[l,:,m_l]\y),
\label{eq:recursive}
\end{align} 
where $\mathbf{u}^{(l-1)} = \sum_{i=1}^{l-1}{{\cal C}_{\mathbf{H}}[i,:,m_i]}$.
Our objective is to find $K$-best candidates with the top-$K$ smallest \textit{score metric} $s^{(l)}$ for each $l$-th layer and prune all the other candidates using a tree structure shown in Fig. \ref{fig:kbest}. We start from the root of the tree and initialize the score $s^{(0)} = 0$. For the $k$-th ($k\in [1,K]$) survived node in the $(l-1)$-th layer with accumulated indices $(m^k_1,\cdots,m^k_{l-1})$, the metrics of all its children nodes with index $m_l \in [1,M]$ are calculated based on \eqref{eq:recursive}, satisfying
\begin{align}
s^{(l)}_{m^k_1,\cdots,m^k_{l-1},m_l} = s^{(l-1)}_{m^k_1,\cdots,m^k_{l-1}}  + ||{\cal C}_{\mathbf{H}}[l,:,m_l]||_2^2
   + 2\Re({\cal C}'_{\mathbf{H}}[l,:,m_l]\mathbf{u}^{(l-1)}_k- {\cal C}'_{\mathbf{H}}[l,:,m_l]\y),
\label{eq:kbest_metric}
\end{align}
where $\mathbf{u}^{(l-1)}_k = \sum_{i=1}^{l-1}{{\cal C}_{\mathbf{H}}[i,:,m^k_i]}$. In this way, $K M$  metrics are obtained and we only preserve the top-$K$ candidates to serve as the survived parent nodes for the next layer whereas all the other candidates are pruned from the tree.  By repeatedly extending and pruning the $K$-best tree, $K$ survived paths are obtained at the last layer. The accumulated indices from the layer 1 to $V$ of the $k$-th survived path are denoted as $(m^k_1,\cdots,m^k_V)$. By converting each $m^k_v$ to a bit sequence $\mathbf{b}^k_v$ and concatenating them together, we obtain a bit sequence $\mathbf{b}^k$ for the subsequent CRC validation.

A well-known weakness of the $K$-best decoding algorithm is the error propagation. Any error made in  previous layers can mislead the decisions in the following layers. To mitigate this issue, we follow the principle in \cite{HDMkbest} to first decode the vectors from ${\cal C}_{\mathbf{H}}$ that are more `reliable' based on the score metric calculated during the tree search by changing the decoding order of remaining layers in the tree. Two different sorting approaches are proposed in \cite{HDMkbest}, namely, \textit{per-layer} sorting and \textit{per-branch} sorting. For  \textit{per-layer} sorting, we  calculate the score metric $s^{(l)}$ assuming each of the remaining $(V-l+1)$ layers as a possible $l$-th layer following \eqref{eq:recursive} and using $\mathbf{u}^{(l-1)}$ of the up-to-now best candidate (with the smallest $s^{(l-1)}$). Then a layer with the minimum score metric is selected as the $l$-th layer to be processed next for all the $K$ survivors.  \textit{Per-branch} sorting also calculates the score metric $s^{(l)}$ of candidates for all remaining layers to determine the order. However, the layer evaluation is specific for each of the $K$ survivors that has a unique accumulated vector $\mathbf{u}_k^{(l-1)}$. As a result, different survivors at each tree level may have distinct decoding orders. Since  \textit{per-branch} sorting determines a specific decoding order for each survivor, it has higher complexity, but it attains superior performance as each survivor can exploit a unique and better ordering for itself in general. 

\subsection{CRC-assisted Looped $K$-best Decoding}
While \textit{per-layer} and \textit{per-branch} sorting approaches improve the error rate performance, any errors made in  previous layers still cannot be corrected in the subsequent layers in the $K$-best algorithm. To address that issue, we propose a \textit{looped} $K$-best decoding algorithm that can correct errors in previously visited layers of the tree to further improve the PER performance.

\begin{figure}[t]
\centering
\includegraphics[width=0.8\linewidth]{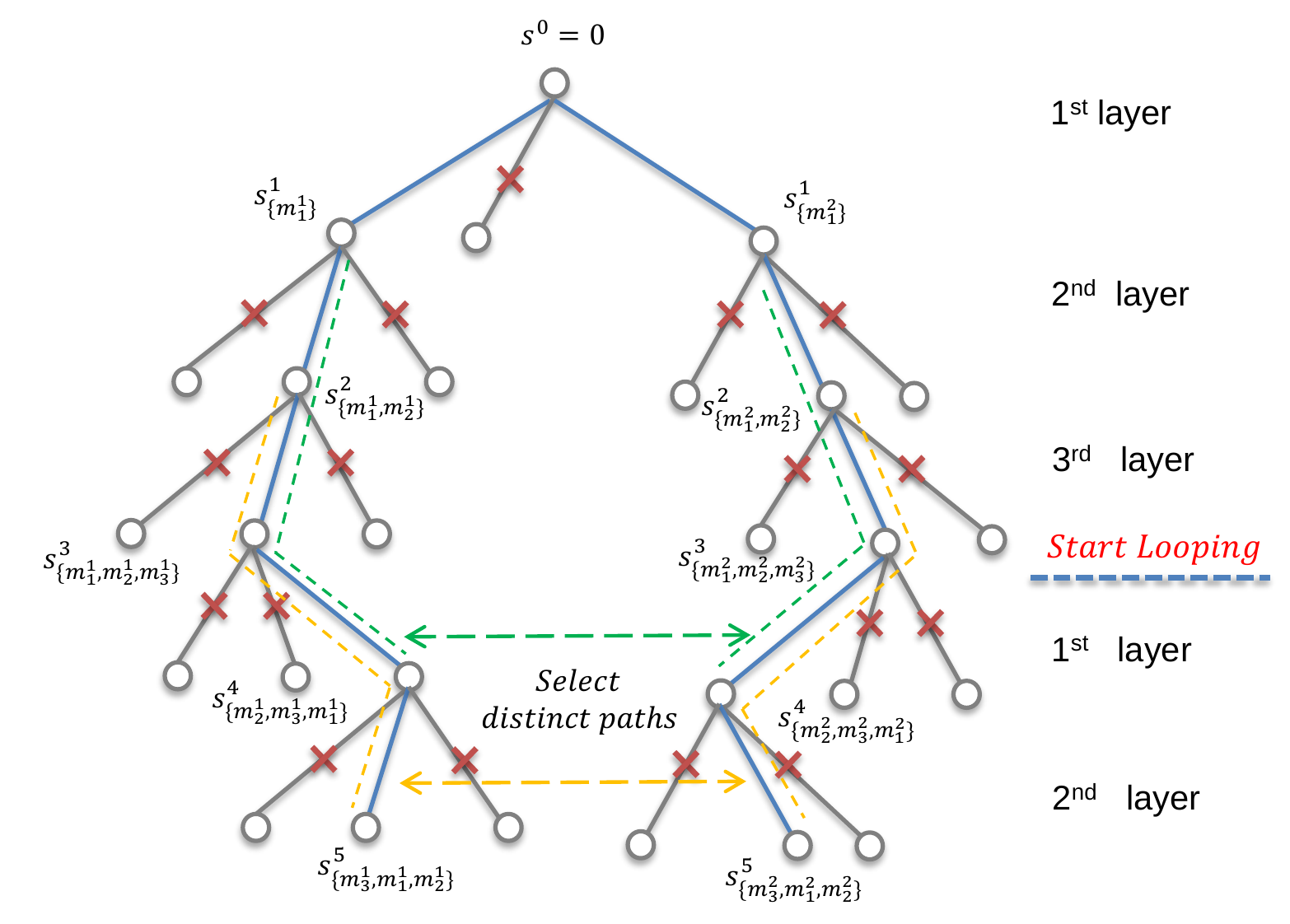}
\caption{The proposed looped $K$-best algorithm with parameter $(V=3, M=3, K=2)$ with two additional iterations. The decoding order $(\pi_1,\pi_2, \pi_3)$ is assumed to be $(1,2,3)$. The two blue branches indicate two survived paths in this example.}
\label{fig:loop_alg}
\end{figure}

The proposed looped $K$-best decoding algorithm performs $Iter$ additional layers of $K$-best decoding to revisit layers that were previously processed. After finishing  regular $K$-best decoding for the final $V$-th layer, $K$ survivors are obtained with corresponding accumulated indices $(m^k_{\pi_1},\cdots,m^k_{\pi_V})$, the score metric $s^{(V)}_{m^k_{\pi_1},\cdots,m^k_{\pi_V}}$, and the decoding order $(\pi_1,...,\pi_V)$\footnote{Although we assume  \textit{per-layer} sorting for simplicity, it is straightforward to extend it to  \textit{per-branch} sorting.}. To proceed to the next additional iteration of $K$-best decoding, it  first updates $K$ score metrics for these survivors by subtracting the terms that correspond to ${\cal C}_{\HM}[\pi_1,:,m_{\pi_1}^k]$ to obtain:
\begin{align}
\tilde{t}^{(V+1)}_{m^k_{\pi_2},\cdots,m^k_{\pi_V}} = s^{(V)}_{m^k_{\pi_1},\cdots,m^k_{\pi_V}}-||{\cal C}_{\mathbf{H}}[\pi_1,:,m^k_{\pi_1}]||_2^2
-2\Re({\cal C}'_{\mathbf{H}}[\pi_1,:,m^k_{\pi_1}]\sum_{v=2}^{V}{{\cal C}_{\mathbf{H}}[\pi_v,:,m^k_{\pi_v}]}-{\cal C}'_{\mathbf{H}}[\pi_1,:,m^k_{\pi_1}]\mathbf{y}),
\label{eq:cancel_metric}
\end{align}
where $\tilde{t}$ denotes the updated metric. Then it repeats the standard process of the (revisited) first layer in the $K$-best decoding algorithm using the survived nodes as the parents by calculating the new score metrics of their children nodes with the index $m_{\pi_1}\in [1,M]$:
\begin{align}
s^{(V+1)}_{m^k_{\pi_2},\cdots,m^k_{\pi_V},m_{\pi_1}} = \tilde{t}^{(V+1)}_{m^k_{\pi_2},\cdots,m^k_{\pi_V}} +||{\cal C}_{\mathbf{H}}[\pi_1,:,m_{\pi_1}]||_2^2
+2\Re({\cal C}'_{\mathbf{H}}[\pi_1,:,m_{\pi_1}]\sum_{v=2}^{V}{{\cal C}_{\mathbf{H}}[\pi_v,:,m^k_{\pi_v}]}-{\cal C}'_{\mathbf{H}}[\pi_1,:,m_{\pi_1}]\mathbf{y}).
\label{eq:update_metric}
\end{align}
One important aspect in the proposed looped $K$-best is that, among the newly generated $KM$ candidates from the revisited layer, it only selects $K$ \textit{distinct} candidates with the best score metrics obtained with the updated accumulated indices $(m^k_{\pi_2},\cdots,m^k_{\pi_V},m^k_{\pi_1})$ and new ordering $(\pi_2,\cdots,\pi_V,\pi_1)$. These indices are reordered to $(m^k_1,\cdots,m^k_V)$ which will be further converted into bit sequences. This process repeats for the next revisited layer until $Iter$ additional layers of $K$-best tree decoding are processed. Fig. \ref{fig:loop_alg} depicts the decoding process of the looped $K$-best decoding algorithm using an example with $(V=3, M=3, K=2)$ and $Iter=2$. 

An interesting property of the proposed looped $K$-best decoding algorithm is that the score metrics of the $K$ survivors are non-increasing with respect to $Iter$. It is expected as we revisit the first element $m^k_{\pi_1}$ for the $k$-th survived path, it is always possible to choose the original element $m^k_{\pi_1}$ selected in the previous round, maintaining the same score metrics. However, in many cases, the algorithm can find new paths with smaller score metrics to improve the performance.

We emphasize that the looped $K$ best needs a new constraint (which is unnecessary in the original $K$-best algorithm) to select \textit{distinct} paths from $KM$ candidates that have unique metrics \eqref{eq:update_metric} without duplication. In the original $K$-best decoding without a loop, the first $K$ survivors from the first layer are always different although they might share the same path for the remaining $(V-1)$ layers. One possible example is $(m^{(1)}_{\pi_1},m_{\pi_2},\cdots,m_{\pi_V})$ and $(m^{(2)}_{\pi_1},m_{\pi_2},\cdots,m_{\pi_V})$ as the final $K=2$ candidates. In this case, when the first branch is revisited during the looped $K$-best decoding, it is likely that these two survived paths select the same $m_{\pi_1}$ making the two paths identical and reducing the effective $K$ from 2 to 1. To avoid such conditions, the proposed algorithm is constrained to only maintain \textit{distinct} survivor paths by eliminating duplicated paths with the same score metric. For that, we first sort the $KM$ score metrics in an increasing order and then eliminate duplicated metrics in the list before we select the final $K$ best unique survivor metrics. 

Once the algorithm finishes processing $Iter$ additional layers, $K$ survived paths (after ordering them back to the original transmit order) are converted to $K$ bit-sequences $\mathbf{b}^k$. Finally, we pass them to check the CRC bits for error detection. A candidates $\mathbf{b}^k$ with a smaller metric is checked first until one that passes the CRC bits is identified as the final decoding output. The entire CRC-assisted looped $K$-best decoding algorithm for the learned MIMO-NOS code is summarized in Algorithm \ref{algorithm}. 
\begin{algorithm}
	\caption{CRC-aided looped $K$-best decoding algorithm with \textit{per-layer} sorting.}
    \label{algorithm}

	\SetKwInOut{Input}{Input}\SetKwInOut{Output}{Output}
	\SetKwFunction{KBest}{KBest}
	\SetKwFunction{CRCcorrection}{CRCcorrection}
	\SetKwFunction{IdxToBits}{IdxToBits}
	\SetKwFunction{CRCDecode}{CRCDecode}
	\SetKwFunction{Reorder}{Reorder}
	\SetKwFunction{ChooseLayer}{ChooseLayer}
	\SetKwFunction{SelectNodes}{SelectNodes}
	\SetKwFunction{SelectDistinctNodes}{SelectDistinctNodes}
	\SetKwData{list}{outputList}
	\SetKwData{idx}{idx}
	\SetKwData{errFlag}{errFlag}
	\SetKwData{decodedBits}{decodedBits}
	\SetKwData{anc}{anc}
	
	\Input{$K, Iter, \mathbf{y}, {\cal C}_{\HM}$}
	\Output{\decodedBits, \errFlag} 
	
	\BlankLine
	\For{$k=1$ \KwTo $K$}{
		$\mathbf{u}(k) \leftarrow 0$ \hfill (zero accumulative vector) \\ 
		$s(k) \leftarrow 0$ \hfill (zero score metric)\\
		$\idx(k) \leftarrow [\;]$ \hfill (empty candidate index)\\
		
	}
    $\mathcal{L} \leftarrow [\;]$ \hfill (empty decoded layer index)\\
	\For{$v=1$ \KwTo $V$}{
	    $l_v \leftarrow$ \ChooseLayer{$\overline{\mathcal{L}}$}\\
	    $\mathcal{L} \leftarrow$ $[\mathcal{L}, l_v]$\\
		\For{$k=1$ \KwTo $K$}{
			
			$\textbf{s}^{tmp}(k) \leftarrow s(k) - 2\Re(\y'{\cal C}_{\HM}[l_v,:,:]-\mathbf{u}'(k){\cal C}_{\HM}[l_v,:,:]) + Diag({\cal C}_{\HM}'[l_v,:,:]{\cal C}_{\HM}[l_v,:,:])$\\
		}
		$[\mathbf{s}, \idx_{new}, \anc] \leftarrow$ \SelectNodes{$\mathbf{s}^{tmp}, K$}\\
		\For{$k=1$ \KwTo $K$}{
			$\mathbf{u}(k) \leftarrow \mathbf{u}(\anc(k)) + {\cal C}_{\HM}[l_v,:,\idx_{new}(k)] $\\
			$\idx(k) \leftarrow [\idx (\anc(k)), \idx_{new}(k)]$\\
		}
	}
	\For{$v=1$ \KwTo $Iter$}{
	    $l_v, \idx_v \leftarrow \mathcal{L}(v), \idx(:,v)$\\
	    $\mathcal{L} \leftarrow$ $[\mathcal{L}, l_v]$\\
		\For{$k=1$ \KwTo $K$}{
			$\mathbf{u}(k), I_{vk} \leftarrow \mathbf{u}(k) - {\cal C}_{\HM}[l_v,:,\idx_v(k)], \idx_v(k)$\\
			$t^{tmp}(k) \leftarrow s(k) + 2\Re(\y'{\cal C}_{\mathbf{H}}[l_v,:,I_{vk}] - \mathbf{u}'(k) {\cal C}_{\mathbf{H}}[l_v,:,I_{vk}]) -||{\cal C}_{\HM}[l_v,:,I_{vk}]||^2_2$  \\
			$\mathbf{s}^{tmp}(k) \leftarrow t^{tmp}(k) -2\Re(\y'{\cal C}_{\mathbf{H}}[l_v,;,:] - \mathbf{u}'(k) {\cal C}_{\mathbf{H}}[l_v,:,:]) + Diag({\cal C}_{\HM}'[l_v,:,:]{\cal C}_{\HM}[l_v,:,:])$\\
		}
		$[\mathbf{s}, \idx_{new}, \anc] \leftarrow$ \SelectDistinctNodes{$\mathbf{s}^\text{tmp}, K$}\\
		\For{$k=1$ \KwTo $K$}{
			$\mathbf{u}(k) \leftarrow \mathbf{u}(\anc(k)) + {\cal C}_{\HM}[l_v,:,\idx_{new}(k)] $\\
			$\idx(k) \leftarrow [\idx (\anc(k)), \idx_{new}(k)]$\\
		}
	}
	$\idx, \mathcal{L} \leftarrow \idx(:,Iter:end), \mathcal{L}(Iter:end)$\\
	\list $\leftarrow$ \Reorder{$\idx$,$\mathcal{L}$}
	
	\While{\errFlag $\neq 0$ and $k \le K$}{
		\decodedBits $\leftarrow$ \IdxToBits{\list$(k)$}\\
		\errFlag $\leftarrow$ \CRCDecode{\decodedBits}\\
	}

	\BlankLine
	\BlankLine
\end{algorithm}

%% file: evaluation.tex
\section{Evaluation}

The PER performance of the proposed scheme is evaluated via Monte-Carlo simulations\footnote{Source code is available at https://github.com/aprilbian/MIMO-NOS}. For short MIMO message transmission, we compare the performance of the learned MIMO-NOS coding using the CRC-assisted looped $K$-best decoding algorithm with a polar-coded MIMO-QPSK (quadrature phase shift keying) scheme demodulated/decoded by maximum-likelihood MIMO detection and CRC-assisted list polar decoding. We also compare the performance of the proposed looped K-best decoding with the neural network-based NOS decoder that is used to train/learn the NOS codebook.

\subsection{Deep Learning Model Training}
The neural network structure shown in Fig. \ref{fig:network} is defined by the parameter set $(V,M,D,N_t,N_r, H_1,H_2)$ where $H_1$ denotes the number of hidden neurons in the encoder $Enc_v$ and decoder $Dec_v$, $v\in [1,V]$, and $H_2$ is the number of hidden neurons in the residual connection module $Res$. We set $H_1=4D, H_2=128$ for all experiments. All DNN models are trained for $5\times 10^3$ epochs with $5\times 10^5$ training samples (packets or codewords) for each epoch. During  training, each training sample/packet observes an independent realization of the random MIMO channel matrix $\mathbf{H} \in \mathbb{C}^{N_r\times N_t}$ as described in Section II. The batch size is set to 1024 and the dynamic learning rate changes linearly from the initial value of $2\times 10^{-4}$ to the final $2\times10^{-6}$. All models are trained under a fixed SNR of $10$dB although they are evaluated under different mismatched SNRs. Once the deep learning model training is complete, we construct a lookup table (LUT) of the learned codebook, $\cal C$ as defined in \eqref{eq:codebook}. 

\subsection{Performance of the Looped $K$-best Decoder}

\begin{figure}[t]
\centering
\includegraphics[width=0.6\linewidth]{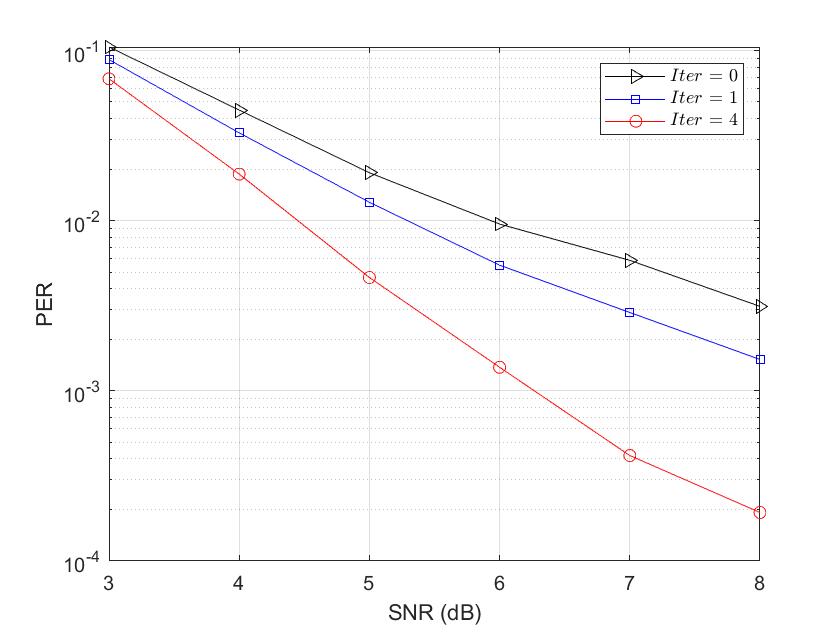}
\caption{PER performance of the CRC-assisted looped $K$-best decoder for the learned MIMO-NOS codebook trained with the system parameter set $(V=4, M=256, D=64, N_t=N_r=4)$. $K=16$ and CRC length is 11 bits. The number of additionally processed layers for the $K$-best decoder is set by $Iter$.}
\label{fig:compare_loop_kbest}
\end{figure}

For PER evaluation, each packet goes through an independent MIMO channel $\mathbf{H}$ while the channel stays the same for a single packet. Fig. \ref{fig:compare_loop_kbest} shows the performance of the CRC-assisted looped $K$-best decoder given the system parameter set of $(V=4, M=256, D=64, N_t=N_r=4)$. This corresponds to transmitting 32 ($=V \cdot log_2(M)$) information bits (including CRC bits) with 4  transmit ($N_t$) and receive ($N_r$) antennas with 8 ($M_c=D/2/N_t$) MIMO channel uses. 
In Fig. \ref{fig:compare_loop_kbest}, $K$ is 16, the CRC length is 11 bits, and $Iter$ denotes the number of additional layer decoding iterations. $Iter=0$ corresponds to the original $K$-best decoding without any loop. Relatively worse performance of $Iter=0$ is expected since the errors made in earlier layers can not be corrected without additional loops. The looped $K$-best algorithm with a higher $Iter$, on the other hand, can correct some previous errors and it attains a 2dB gain with $Iter=4$ for PER $\approx 10^{-2}$. 

\begin{figure}
\centering
\includegraphics[width=0.6\linewidth]{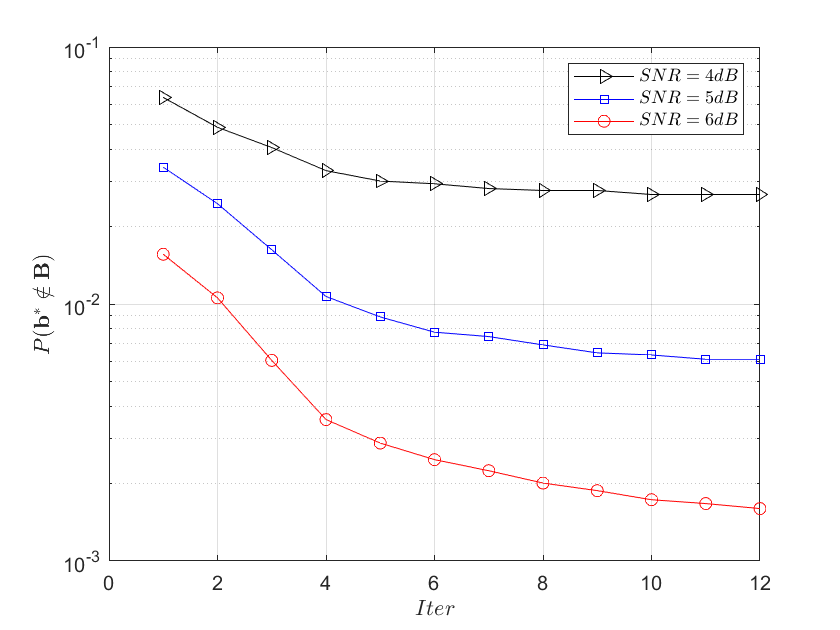}
\caption{Error rate performance the looped $K$-best algorithm with various $Iter$ settings given the parameter set $(V=6,M=256,D=96,N_t=N_r=4)$ under different SNRs.}
\label{fig:loop_kbest}
\end{figure}

\begin{figure}[t]
\centering
\includegraphics[width=0.6\linewidth]{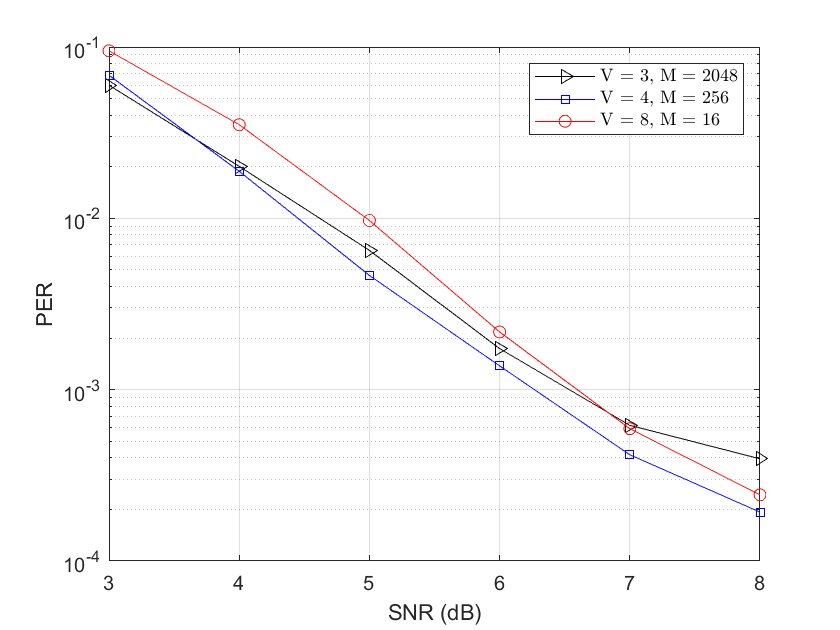}
\caption{PER performance of different $(V,M)$ combinations with $\approx 32$ information bits for a $4\times 4$ MIMO system. There is an optimal $V$ for the target rate as the performance is not a monotonic function of $V$ given the target rate.}
\label{fig:VM_PER}
\end{figure}

We then evaluate the error rate performance of the looped $K$-best algorithm with respect to a wide range of $Iter$ in Fig. \ref{fig:loop_kbest} for MIMO-NOS code trained with the parameter set $(V=6,M=256,D=96,N_t=N_r=4)$ and evaluated at different SNRs. $K=16$ and the CRC length is 11 bits. The error rate performance is quantified using the probability $P(\mathbf{b}^* \notin \mathbf{B})$ where $\mathbf{b}^*$ is the correct bit sequence and $\mathbf{B}$ is the set of  $K$-best candidates $\mathbf{b}^k, k\in [1,K]$, obtained by the algorithm. As $Iter$ increases, $P(\mathbf{b}^* \notin \mathbf{B})$ monotonically decreases resulting in the improved PER. Figure \ref{fig:loop_kbest} further shows that the error rate performance improvement from the increased number of iterations is more substantial when the SNR is higher. It is observed that the PER stops significantly improving when $Iter \ge V$ in general. Thus, we set $Iter=V$ for the remaining evaluations (unless noted otherwise) to strike a balance between the PER performance and the decoding complexity.

\subsection{Performance with Different System Parameters}

Given the number of transmit antennas $N_t$,  the rate $R$ of the proposed MIMO-NOS scheme is determined by the number of information bits ($V \cdot log_2(M)$) and the length of the complex-valued codeword ($D/2$), satisfying $R = \frac{N_t V log_2(M)}{D/2}$.  With a fixed $D$, there are different $(V, M)$ combinations to obtain the same target rate $R$ whereas one configuration outperforms the other. Our prior work \cite{NOS} for single antenna AWGN channel argues that the number of superimposed vectors $V$ should be minimized (with a larger $M$) as long as the complexity (i.e., model size) of the neural network to learn an NOS codebook is manageable. However, we find that for the proposed MIMO-NOS coding, using a smaller $V$ (and larger $M$) does not necessarily improve the PER performance while it definitely increases the complexity of the network model. The analysis is involved but numerical evaluation of the score metric in \eqref{eq:kbest_metric} under the MIMO channel shows that there is an optimal $V$ (and corresponding $M$) that balances the inter- and intra-codeword correlation tradeoff. Fig. \ref{fig:VM_PER} shows the PER performance of three different $(V,M)$ combinations that are $(V=3,M=2048)$, $(V=4,M=256)$ and $(V=8,M=16)$ evaluated under $4 \times 4$ MIMO transmission with $D=64$, $K=16$, and 11-bit CRC. Note that all these settings have (almost) the same rate. The setting of $(V=4,M=256)$ outperforms the other with smaller or larger $V$'s. For a fair comparison, $Iter$ is set to 4 for both $(V=3,M=2048)$ and $(V=4,M=256)$ settings  while $Iter=V=8$ is used for $(V=8,M=16)$. We observed that $(V=4,M=256)$ outperforms the other settings when all use unlimited $Iter$.
It is worth noting that the $(V=8,M=16)$ setting is inferior to  $(V=3,M=2048)$ at low SNRs while the opposite is observed at high ($>$7dB) SNRs. It is because of the tradeoff between inter- and intra-codeword distances that the proposed $K$-best algorithm experiences during the decoding process. A larger $V$ (smaller $M$) creates more severe inter-codeword interference with a deeper tree structure that makes the algorithm suffer from early decoding errors in the tree at low SNRs. When the SNR is relatively high with lower chance of early stage errors in the $K$-best decoding, the performance is limited by the intra-codeword distance as more candidates $M$ are evaluated for each layer. Although it is difficult to accurately analyze this tradeoff, Fig. \ref{fig:VM_PER} shows that there is an optimal parameter set and the PER performance is not necessarily a monotonic function of $V$ or $M$. Empirically, we observed that a setting with $M=256$ usually outperforms others (as observed in Fig. \ref{fig:VM_PER}). Hence, we use $M=256$ (with a corresponding $V$ to attain the target rate) for the rest of the paper to evaluate the performance of the proposed MIMO NOS scheme.

In the proposed scheme, the dimension of the codebook ${\cal C}$ \eqref{eq:codebook} is determined by the parameter set $(V,M,D)$ and it does not depend on the MIMO configuration, $(N_t, N_r)$. For a given codebook ${\cal C}$, the MIMO configuration $(N_t, N_r)$ defines the space time coding scheme by reshaping the samples of a transmitted codeword with proper space and time indices as discussed in Secion III.B. This implies that it is possible to use a codebook for different MIMO settings by simple reshaping even though they are not necessarily identical to that used during the codebook training. In other words, one can apply reshaping based on the desired $(N_t, N_r)$ to an existing learned codebook trained with different $(N_t, N_r)$ as long as $(V,M,D)$ is unchanged. 
To facilitate the discussion to follow, we distinguish the number of transmit and receive antennas used during the training by $N_t^l$ and $N_r^l$, respectively. Consequently, $N_t$ and $N_r$ denotes the number of antennas for evaluation of a learned MIMO NOS codebook. We observed that $N_r^l$ makes little impact to the PER performance of the codebook for a given evaluation setup $N_t$ or $N_r$ as long as $N_r^l \geq N_t^l$ holds. Thus we only show the impact of $N_t^l(=N_r^l)$ in the following discussion.

\begin{figure}[t]
\centering
\includegraphics[width=0.6\linewidth]{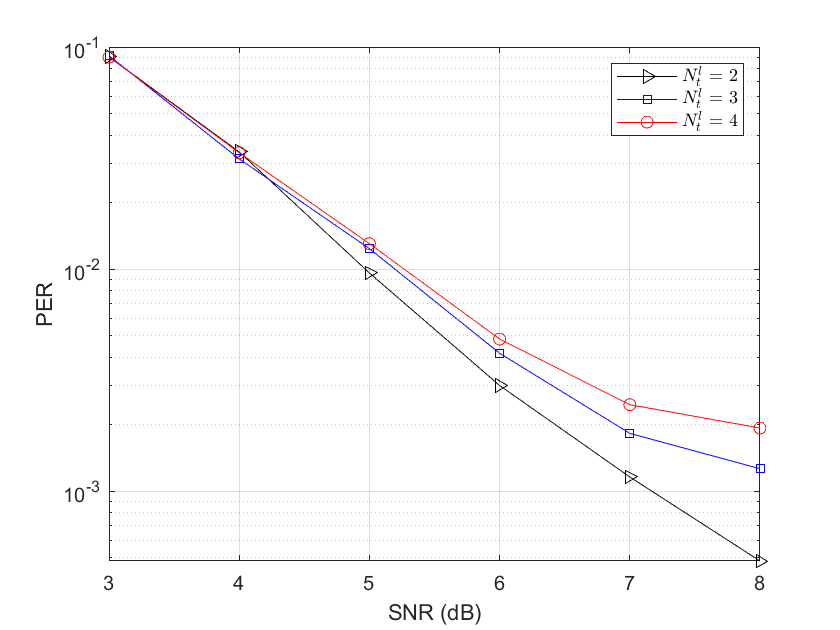}
\caption{The PER performance of the codebooks learned under $N_t^l=2,3,4$ applied to the $4 \times 4$ MIMO transmissions with parameters $(V=6,M=256,D=96,K=16,Iter=6,CRCLen=11)$}
\label{fig:per_codebook}
\end{figure}

Fig. \ref{fig:per_codebook} shows the PER performance of the codebooks for the setting $(V=6,M=256,D=96)$ trained with $N_t^l=N_r^l=2,3,4$ and evaluated for $N_t = N_r = 4$ MIMO transmission. We set $K=16$ and the CRC length is 11 bits. Intuitively, one would expect the best performance when $N_t^l=N_t$. However, the simulation shows that the codebooks trained with $N_t^l = 2$ or $3$  outperform the one with $N_t^l=4$ for the $N_t=4$ evaluation, showing the `mismatch' between $N_t^l$ and $N_t$ for the optimal performance. 

\begin{figure}[t]
    \centering
	\begin{subfigure}{0.6\linewidth}
		\centering
		\includegraphics[width=0.9\linewidth]{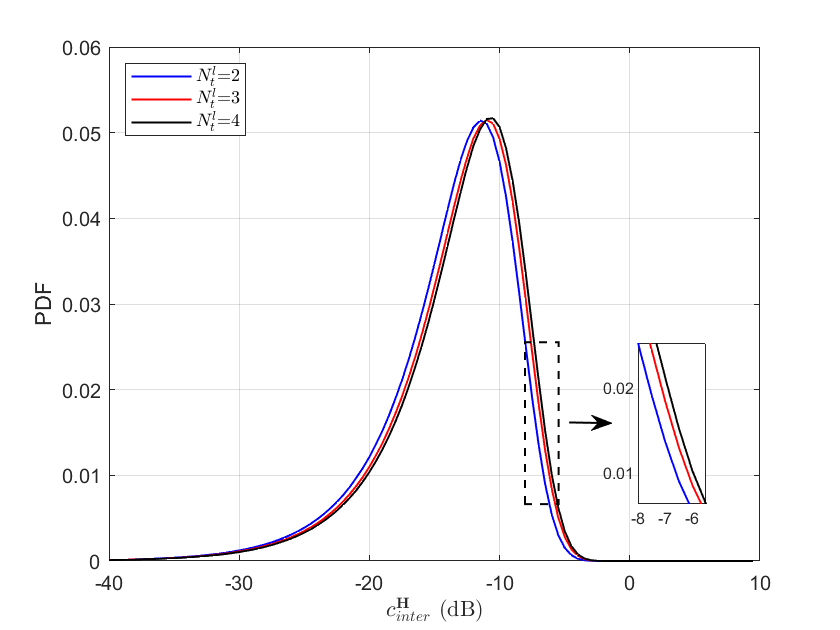}
		\caption{\textit{Inter}-correlation distribution}
		\label{inter}
	\end{subfigure}

	\vspace{4mm}
	\begin{subfigure}{0.6\linewidth}
		\centering
		\includegraphics[width=0.9\linewidth]{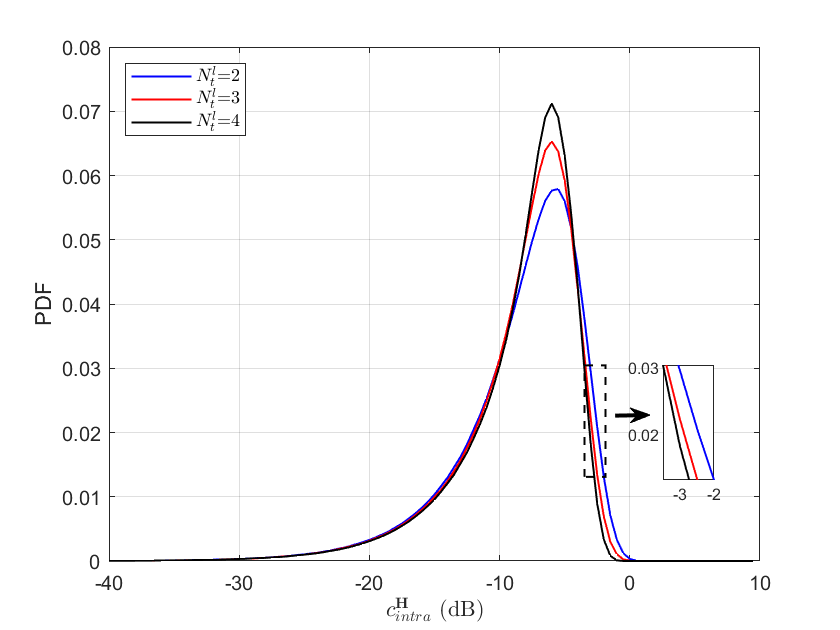}
		\caption{\textit{Intra}-correlation distribution}
		\label{intra}
	\end{subfigure}
	\vspace{2mm}
	\caption{\textit{Inter}- and \textit{Intra}-correlation distributions for codebooks learned under different transmit antennas $N_t^l=2,3,4$ evaluated for $4\times 4$ MIMO transmission given the system parameter $(V=6,M=256,D=96)$.}
	\label{diff_codebook}
\end{figure}

To understand this mismatch, Fig. \ref{diff_codebook} analyzes inter-correlation $c_{inter}^{\HM}$ and intra-correlation $c_{intra}^{\HM}$ for different $N_t^l$'s with random $4\times4$ MIMO channel realizations. Notice that the codebooks trained with $N_t^l=$2 or 3 have better $c_{inter}^{\HM}$ distribution compared with the $N_t^l=4$ counterpart, while the $N_t^l=4$ codebook has better $c_{intra}^{\HM}$ distribution. From this experiment using the given parameter set, we observe that the PER of the proposed looped $K$-best decoding is dominated by the inter-codeword interference that propagates down to later tree levels during the looped $K$-best decoding. When inter-codeword interference is correctly cancelled out, decoding of each layer (whose performance is governed by intra-codeword correlation) using a reasonably high $K\gg 1$ with respect to $M$ does not limit the PER performance at high SNRs. The codebook learned with $N_t^l=2$ strikes the balance between inter- and intra-codeword correlation for $N_t=4$ evaluation. 


The above observation brings one question why the proposed MIMO NOS framework learns a better codebook under a mismatched MIMO scenario $N_t^l \neq N_t$. It can be explained by the mismatch between the  hand-crafted looped $K$-best decoder used for evaluation and the neural network-based one-shot decoder used for training as introduced in Section II. Although the looped $K$-best coding outperforms the neural network based decoder (as shown in the next subsection), it is not differentiable and thus cannot be directly used as a decoder for the end-to-end training to learn a codebook. Since the training is performed with a sub-optimal neural network-based decoder, the property of learned codebook is not necessarily optimal for the proposed looped $K$-best coding algorithm. This mismatch can potentially be resolved by approximating the $K$-best algorithm to a differentiable method for end-to-end training (which is left as future work).


\begin{figure}
    \centering
	\begin{subfigure}{0.6\linewidth}
		\centering
		\includegraphics[width=0.9\linewidth]{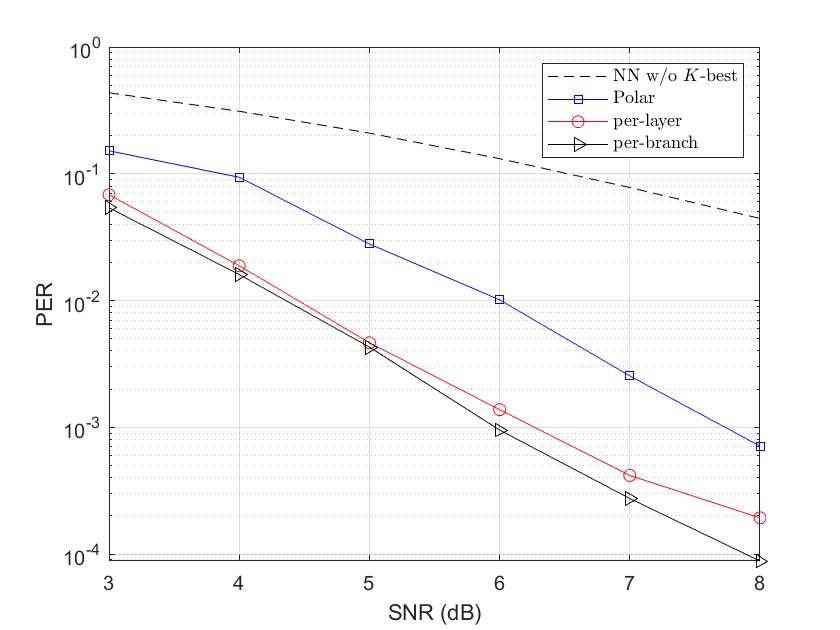}
		\caption{32 info bits, $(V=4,M=256,D=64)$}
	\end{subfigure}
    \vspace{3mm}

	\begin{subfigure}{0.6\linewidth}
		\centering
		\includegraphics[width=0.9\linewidth]{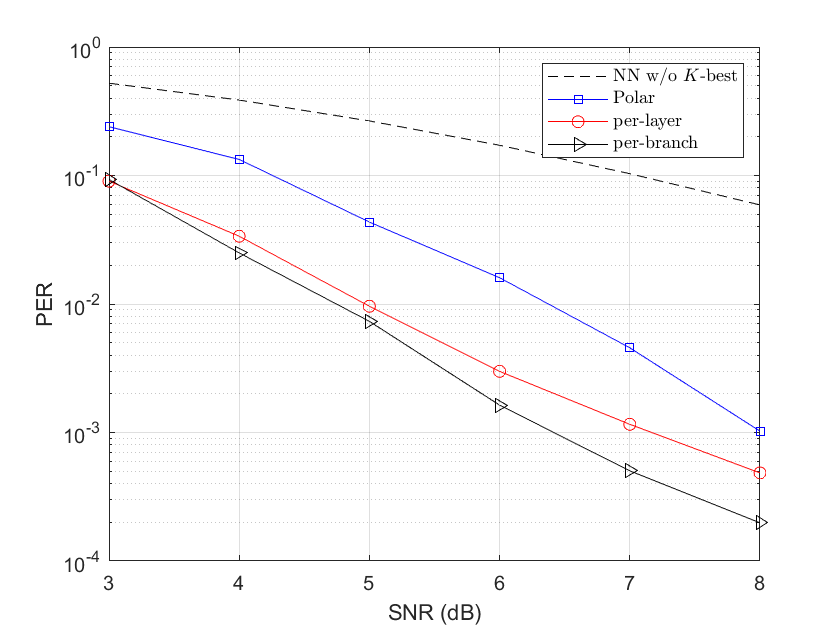}
		\caption{48 info bits, $(V=6,M=256,D=96)$}
	\end{subfigure}
    \vspace{3mm}

	\begin{subfigure}{0.6\linewidth}
		\centering
		\includegraphics[width=0.9\linewidth]{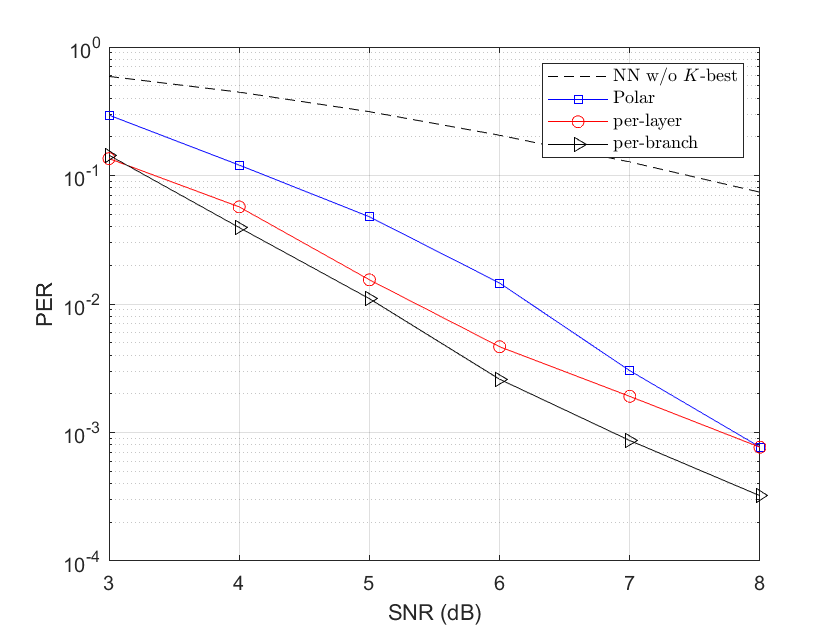}
		\caption{64 info bits, $(V=8,M=256,D=128)$}
	\end{subfigure}
	\vspace{2mm}
	
	\caption{The proposed MIMO-NOS scheme outperforms the polar with ML MIMO detection under different number of information bits ranging from 32 to 64 bits in 4$\times$4 MIMO channels.}
	\label{fig:final_plot}
\end{figure}

\subsection{PER Performance Comparison with a Conventional Scheme}
Finally, we compare the performance of our MIMO-NOS scheme with the conventional polar-coded MIMO system. As discussed in \cite{polarld}, the CRC-assisted polar code is proven to be robust for short packet transmission, thus we selected it as the baseline. Although there exist multiple computationally-efficient MIMO detection algorithms such as sphere decoding and $K$-best decoding\cite{Sphere_Dec,kbest2} that provide soft decisions, we choose the ML MIMO detection for the baseline to avoid degrading the polar code performance. We apply a successive cancellation list decoding algorithm (SCL) \cite{polarld} with list size $L$ to polar decoding.

For comparisons with CRC-assisted list polar decoding, we train the MIMO-NOS codebook with $N_T^l=N_R^l=2$, and evaluate both in the $4\times 4$ MIMO configuration. In the first case, we evaluate transmittion of 32 message bits (Fig. \ref{fig:final_plot} (a)), and in the second and third case we increase the message length to 48 bits (Fig. \ref{fig:final_plot} (b)) and 64 bits (Fig. \ref{fig:final_plot} (c)). The MIMO-NOS scheme uses the parameter set $(V = 4, M = 256, D = 64)$ for the first case (32 bits), $(V=6,M=256,D=96)$ for the second (48-bit), and $(V=8,M=256,D=128)$ for the last case (64-bit) while $K=16$ for all these cases.  The baseline uses 3GPP polar code \cite{3GPP_polar} with QPSK modulation and $0.5$ coding rate for all these cases. Its list decoding size $L$ is set to be $L=K=16$ for fair comparison. The 11-bit CRC with a generator polynomial $x^{11}+x^{10}+x^9+x^5+1$ is adopted to both MIMO-NOS and the polar code baseline. Note that all these schemes have the same spectral efficiency of $4$ bits/Hz/sec with $N_t=4$. 

Fig. \ref{fig:final_plot} shows the MIMO-NOS scheme (with either \textit{per-branch} or \textit{per-layer}  sorting) outperforms the polar baseline by $1-2 dB$ for short messages in the range of 32 -- 64 bits. The `NN w/o $K$-best' curve shows the PER performance of the neural decoder (without the aid of CRC bits) introduced in Section II and used for codebook training. When it is trained and tested with the residual connection network $Res$ in parallel with the conventional MMSE detection in the setting of $N_T^l = N_T = 2$, the performance of the neural network-based decoder improves by $\approx2$dB compared to a version without $Res$ connection. However, the neural network decoder using $Res$ connection still turns out to be significantly inferior to the proposed looped $K$-best decoder (when both are evaluated without CRC bits). The SNR gain of the \textit{per-branch} sorting over the \textit{per-layer} sorting improves with the number of information bits from approximately $0.2 dB$ for 32 bits to  $1 dB$ for 64 bits. Fig. \ref{fig:final_plot} (c) shows that the SNR gain of the proposed scheme over the polar baseline reduces with a larger number of information bits, which is expected because polar coding is capacity achieving when the codeword length is sufficiently long.  

\begin{figure}
\centering
\includegraphics[width=0.6\linewidth]{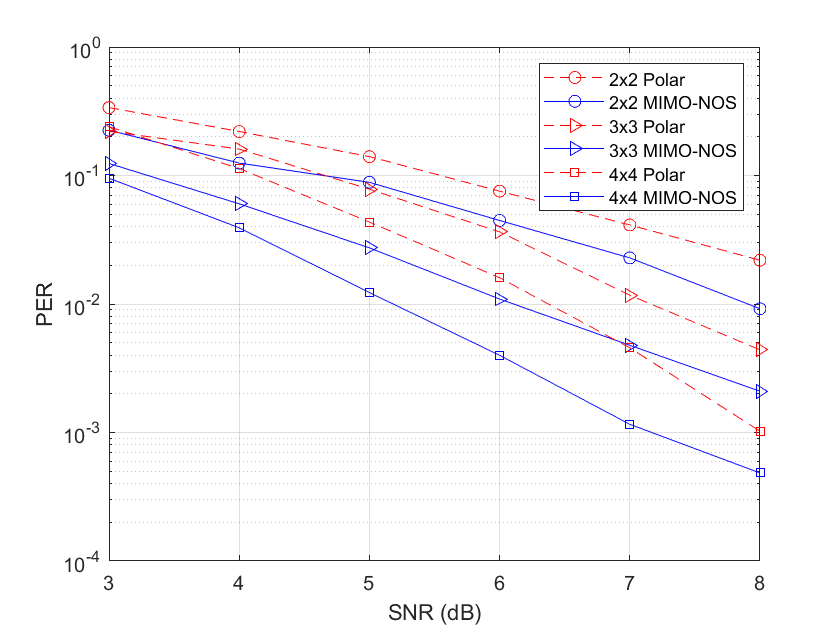}
\caption{The comparison of the MIMO-NOS and the baseline polar code applied to $2\times 2, 3\times 3$ and $4\times 4$  MIMO transmissions given the parameters $(V=6,M=256,D=96)$, and $K=L=16$. CRC length is 11-bit, information length is 48-bits, and the spectral efficiency is $N_T$ bits/Hz/sec.}
\label{fig:mimo_simu}
\end{figure}

We now compare the performance of the MIMO-NOS scheme and the baseline under different MIMO settings. The parameters of the MIMO-NOS for this simulation are $(V=6,M=256,D=96)$, 48 message bits, and $K=16$ with \textit{per-layer} sorting. The codebook is learned with $N_t^l=2$. The SCL  decoding based polar scheme has the same information bit length and coding rate of $0.5$ with QPSK modulation and $L=16$. Both schemes adopt 11-bit CRC and are tested with $2\times 2$, $3\times 3$, and $4\times 4$ MIMO  configurations. Fig. \ref{fig:mimo_simu} shows the proposed MIMO-NOS scheme outperforms the polar baseline for all tested MIMO settings.

\subsection{Discussion}
As shown in Fig. \ref{fig:final_plot}, the performance gap between the learned MIMO-NOS and polar baseline reduces as the message length increases. The complexity of the proposed MIMO-NOS encoder shown in Fig. \ref{fig:network} grows exponentially with the message length (it is proportional to $M$ while the message length is given by $log_2(M)$), thus it is not practical to scale the proposed scheme to an arbitrarily long length although conventional superposition codes are known to be capacity achieving for long sequences \cite{SPARC2014}. Nevertheless, the proposed MIMO-NOS is a promising solution for reliable short message MIMO transmission in the low SNR regime with superior PER performance and an efficient decoding algorithm. Investigating new network structures and corresponding training schemes for learned superposition coding that scales better to longer information bit lengths is left as future work.

%% file: main.bbl
\begin{thebibliography}{10}
\providecommand{\url}[1]{#1}
\csname url@samestyle\endcsname
\providecommand{\newblock}{\relax}
\providecommand{\bibinfo}[2]{#2}
\providecommand{\BIBentrySTDinterwordspacing}{\spaceskip=0pt\relax}
\providecommand{\BIBentryALTinterwordstretchfactor}{4}
\providecommand{\BIBentryALTinterwordspacing}{\spaceskip=\fontdimen2\font plus
\BIBentryALTinterwordstretchfactor\fontdimen3\font minus
  \fontdimen4\font\relax}
\providecommand{\BIBforeignlanguage}[2]{{%
\expandafter\ifx\csname l@#1\endcsname\relax
\typeout{** WARNING: IEEEtran.bst: No hyphenation pattern has been}%
\typeout{** loaded for the language `#1'. Using the pattern for}%
\typeout{** the default language instead.}%
\else
\language=\csname l@#1\endcsname
\fi
#2}}
\providecommand{\BIBdecl}{\relax}
\BIBdecl

\bibitem{NOS}
\BIBentryALTinterwordspacing
C.~Bian, M.~Yang, C.~Hsu, and H.~Kim, ``Deep learning based near-orthogonal
  superposition code for short message transmission,'' \emph{CoRR}, vol.
  abs/2111.03263, 2021. [Online]. Available:
  \url{https://arxiv.org/abs/2111.03263}
\BIBentrySTDinterwordspacing

\bibitem{mMTC1}
Z.~{Dawy}, W.~{Saad}, A.~{Ghosh}, J.~G. {Andrews}, and E.~{Yaacoub}, ``{Toward
  Massive Machine Type Cellular Communications},'' \emph{IEEE Wireless
  Communications}, vol.~24, pp. 120--128, 2017.

\bibitem{mMTC2}
C.~{Bockelmann}, N.~{Pratas}, H.~{Nikopour}, K.~{Au}, T.~{Svensson},
  C.~{Stefanovic}, P.~{Popovski}, and A.~{Dekorsy}, ``{Massive machine-type
  communications in 5g: physical and MAC-layer solutions},'' \emph{{IEEE}
  Commun. Mag.}, vol.~54, no.~9, pp. 59--65, 2016.

\bibitem{mMTC3}
M.~R. Palattella, M.~Dohler, A.~Grieco, G.~Rizzo, J.~Torsner, T.~Engel, and
  L.~Ladid, ``Internet of things in the 5g era: Enablers, architecture, and
  business models,'' \emph{IEEE Journal on Selected Areas in Communications},
  vol.~34, no.~3, pp. 510--527, 2016.

\bibitem{polarld}
I.~{Tal} and A.~{Vardy}, ``List decoding of polar codes,'' \emph{IEEE Trans. on
  Info. Theory}, vol.~61, no.~5, pp. 2213--2226, 2015.

\bibitem{compare_codes}
H.~{Gamage}, N.~{Rajatheva}, and M.~{Latva-aho}, ``Channel coding for enhanced
  mobile broadband communication in 5g systems,'' in \emph{2017 European Conf.
  on Networks and Comm. (EuCNC)}, 2017, pp. 1--6.

\bibitem{Coskun2019}
M.~C. Coşkun, G.~Durisi, T.~Jerkovits, G.~Liva, W.~Ryan, B.~Stein, and
  F.~Steiner, ``Efficient error-correcting codes in the short blocklength
  regime,'' \emph{Physical Communication}, vol.~34, pp. 66 -- 79, 2019.

\bibitem{HDM2018}
H.~{Kim}, ``{HDM: Hyper-Dimensional Modulation for Robust Low-Power
  Communications},'' in \emph{2018 IEEE International Conference on
  Communications (ICC)}, May 2018, pp. 1--6.

\bibitem{HDM2019}
C.~{Hsu} and H.~{Kim}, ``{Collision-Tolerant Narrowband Communication Using
  Non-Orthogonal Modulation and Multiple Access},'' in \emph{2019 IEEE Global
  Communications Conference (GLOBECOM)}, 2019, pp. 1--6.

\bibitem{HDMkbest}
------, ``{Non-Orthogonal Modulation for Short Packets in Massive Machine type
  communication},'' in \emph{2020 IEEE Global Communications Conference
  (GLOBECOM)}, 2020, pp. 1--6.

\bibitem{SPARC2014}
A.~{Joseph} and A.~R. {Barron}, ``{Fast Sparse Superposition Codes Have Near
  Exponential Error Probability for $R < {\cal C}$},'' \emph{{IEEE} Trans. Inf.
  Theory}, vol.~60, no.~2, pp. 919--942, 2014.

\bibitem{deepdec}
T.~{Gruber}, S.~{Cammerer}, J.~{Hoydis}, and S.~t.~{Brink}, ``On deep
  learning-based channel decoding,'' in \emph{2017 51st Annual Conference on
  Information Sciences and Systems (CISS)}, 2017, pp. 1--6.

\bibitem{deepldpc}
E.~{Nachmani}, Y.~{Be'ery}, and D.~{Burshtein}, ``Learning to decode linear
  codes using deep learning,'' in \emph{2016 54th Annual Allerton Conf. on
  Comm., Control, and Computing (Allerton)}, 2016, pp. 341--346.

\bibitem{deepae}
T.~{O’Shea} and J.~{Hoydis}, ``An introduction to deep learning for the
  physical layer,'' \emph{IEEE Transactions on Cognitive Communications and
  Networking}, vol.~3, no.~4, pp. 563--575, 2017.

\bibitem{LEARN}
Y.~Jiang, H.~Kim, H.~Asnani, S.~Kannan, S.~Oh, and P.~Viswanath, ``Learn codes:
  Inventing low-latency codes via recurrent neural networks,'' \emph{IEEE
  Journal on Selected Areas in Information Theory}, vol.~1, no.~1, pp.
  207--216, 2020.

\bibitem{jiang2019turbo}
------, ``Turbo autoencoder: Deep learning based channel codes for
  point-to-point communication channels,'' in \emph{Advances in Neural
  Information Processing Systems}, 2019, pp. 2754--2764.

\bibitem{DetNet}
N.~Samuel, T.~Diskin, and A.~Wiesel, ``Deep mimo detection,'' in \emph{2017
  IEEE 18th International Workshop on Signal Processing Advances in Wireless
  Communications (SPAWC)}, 2017, pp. 1--5.

\bibitem{JMIMODD}
T.~Wang, L.~Zhang, and S.~C. Liew, ``Deep learning for joint mimo detection and
  channel decoding,'' in \emph{2019 IEEE 30th Annual International Symposium on
  Personal, Indoor and Mobile Radio Communications (PIMRC)}, 2019, pp. 1--7.

\bibitem{SPARC_AMP}
C.~Rush, A.~Greig, and R.~Venkataramanan, ``Capacity-achieving sparse
  superposition codes via approximate message passing decoding,'' \emph{IEEE
  Trans. on Info. Theory}, vol.~63, no.~3, pp. 1476--1500, 2017.

\bibitem{Finite_SPARC}
A.~Greig and R.~Venkataramanan, ``Techniques for improving the finite length
  performance of sparse superposition codes,'' \emph{IEEE Transactions on
  Communications}, vol.~66, no.~3, pp. 905--917, 2018.

\bibitem{STT}
B.~Hassibi and B.~Hochwald, ``High-rate codes that are linear in space and
  time,'' \emph{IEEE Transactions on Information Theory}, vol.~48, no.~7, pp.
  1804--1824, 2002.

\bibitem{STBC}
V.~Tarokh, H.~Jafarkhani, and A.~Calderbank, ``Space-time block codes from
  orthogonal designs,'' \emph{IEEE Transactions on Information Theory},
  vol.~45, no.~5, pp. 1456--1467, 1999.

\bibitem{MIMO_CHEST1}
M.~Biguesh and A.~Gershman, ``Training-based mimo channel estimation: a study
  of estimator tradeoffs and optimal training signals,'' \emph{IEEE
  Transactions on Signal Processing}, vol.~54, no.~3, pp. 884--893, 2006.

\bibitem{MIMO_CHEST2}
Y.~Li, ``Simplified channel estimation for ofdm systems with multiple transmit
  antennas,'' \emph{IEEE Transactions on Wireless Communications}, vol.~1,
  no.~1, pp. 67--75, 2002.

\bibitem{MIMO_CHEST3}
M.~K. Ozdemir and H.~Arslan, ``Channel estimation for wireless ofdm systems,''
  \emph{IEEE Communications Surveys Tutorials}, vol.~9, no.~2, pp. 18--48,
  2007.

\bibitem{JSCC_OFDM}
M.~Yang, C.~Bian, and H.-S. Kim, ``Ofdm-guided deep joint source channel coding
  for wireless multipath fading channels,'' \emph{IEEE Transactions on
  Cognitive Communications and Networking}, vol.~8, no.~2, pp. 584--599, 2022.

\bibitem{kbest}
{Zhan Guo} and P.~{Nilsson}, ``{Algorithm and implementation of the K-best
  sphere decoding for MIMO detection},'' \emph{{IEEE} J. Sel. Areas Commun.},
  vol.~24, no.~3, pp. 491--503, 2006.

\bibitem{kbest2}
L.~G. {Barbero} and J.~S. {Thompson}, ``Fixing the complexity of the sphere
  decoder for mimo detection,'' \emph{IEEE Trans. on Wireless Comm.}, vol.~7,
  no.~6, pp. 2131--2142, 2008.

\bibitem{Sphere_Dec}
E.~Agrell, T.~Eriksson, A.~Vardy, and K.~Zeger, ``Closest point search in
  lattices,'' \emph{IEEE Transactions on Information Theory}, vol.~48, no.~8,
  pp. 2201--2214, 2002.

\bibitem{3GPP_polar}
G.~T. 38.212, ``{NR; Multiplexing and channel coding (Release 15)},'' \emph{3rd
  Generation Partnership Project}, 2018.

\end{thebibliography}
